\documentclass[preprints,article,accept,moreauthors,pdftex,10pt,a4paper]{Definitions/mdpi}

\firstpage{1} 
\pubvolume{xx}
\issuenum{1}
\articlenumber{5}
\pubyear{2018}
\copyrightyear{2018}
\history{Received: date; Accepted: date; Published: date}

\newcommand{\lb}{\left(}
\newcommand{\rb}{\right)}
\newcommand{\lbc}{\left\{}
\newcommand{\rbc}{\right\}}
\newcommand{\lbq}{\left[}
\newcommand{\rbq}{\right]}

\usepackage{graphicx}
\graphicspath{ {Images/} }
\usepackage{amsmath}

\Title{$\mathcal{S}$-Matrix of Nonlocal Scalar Quantum Field Theory in the Representation of Basis Functions}

\Author{Ivan V. Chebotarev $^{1}$, Vladislav A. Guskov $^{1}$, Stanislav L. Ogarkov $^{1,2\dagger}$ and Matthew Bernard $^{1\ddagger}$}

\AuthorNames{Ivan V. Chebotarev, Vladislav A. Guskov, Stanislav L. Ogarkov and Matthew Bernard}

\address{%
$^{1}$ \quad Moscow Institute of Physics and Technology (MIPT), Institutskiy Pereulok 9, 141701 Dolgoprudny, Moscow Region, Russia\\
$^{2}$ \quad Dukhov Research Institute of Automatics (VNIIA), Sushchevskaya 22, 127055 Moscow, Russia}

\corres{Correspondence: $^{\dagger}$ogarkovstas@mail.ru, $^{\ddagger}$mattb@berkeley.edu}

\abstract{Nonlocal quantum theory of one-component scalar field in $D$-dimensional Euclidean spacetime is studied in representations of $\mathcal{S}$-matrix theory for both polynomial and nonpolynomial interaction Lagrangians. The theory is formulated on coupling constant $g$ in the form of an infrared smooth function of argument $x$ for space without boundary. Nonlocality is given by evolution of Gaussian propagator for the local free theory with ultraviolet form factors depending on ultraviolet length parameter $l$. By representation of the $\mathcal{S}$-matrix in terms of abstract functional integral over primary scalar field, the $\mathcal{S}$ form of a grand canonical partition function is found. And, by expression of $\mathcal{S}$-matrix in terms of the partition function, the representation for $\mathcal{S}$ in terms of basis functions is obtained. Derivations are given for discrete case where basis functions are Hermite functions, and for continuous case where basis functions are trigonometric functions. The obtained expressions for the $\mathcal{S}$-matrix are investigated within the framework of variational principle based on Jensen inequality. And, by the latter, the majorant of $\mathcal{S}$ (more precisely, of $-\ln{\mathcal{S}}$) is constructed. Equations with separable kernels satisfied by variational function $q$ are found and solved, yielding results for both the polynomial theory $\varphi^{4}$ (with suggestions for  $\varphi^{6}$) and the nonpolynomial sine-Gordon theory. A new definition of the $\mathcal{S}$-matrix is proposed to solve additional divergences which arise in application of Jensen inequality for the continuous case. Analytical results are obtained and illustrated numerically, with plots of variational functions $q$ and corresponding majorants for the $\mathcal{S}$-matrices of the theory. For simplicity of numerical calculation: the $D=1$ case is considered, and propagator for the free theory $G$ is in the form of Gaussian function typically in the Virton-Quark model, although the obtained analytical inferences are not limited to these particular choices in principle. The formulation for nonlocal QFT in momentum $k$ space of extra dimensions with subsequent compactification into physical spacetime is discussed, alongside the compactification process.}


\begin{document}

\section{Introduction}

Timeline of Quantum Field Theory (QFT) offers events that are quite asymmetric to each other; brilliant triumphs, on the one hand, in explanations and predictions of different processes at low-energy Quantum Electrodynamics (QED), high-energy Quantum Chromodynamics (QCD), Theory of Critical Phenomena, and a number of other branches of the modern science; but, catastrophe, on the other hand, in various attempts to describe high-energy physics at scalar theory and QED levels as well as low-energy QCD physics. Identically, in discoveries of renormalizable field theories, the conjecture that only those make sense is opposing to conjecture that there are any and all theories. Nonetheless, the robust indeterminacy is fueled by non-speculative mathematically theory.

In Non-Abelian Gauge QFT, unifying electromagnetic and weak interactions into one general model, weak-interaction processes are consistently described. In another triumph of Non-Abelian QFT, the Standard Model (SM), high-energy problem of QED vanishes in chromodynamic reaction channels, because QCD consistently at high-energy region has the quintessential property of asymptotic freedom. Recently, the discovery of Higgs boson, the last SM element in energy domain, where existence is most natural, occurred: Notably, by the remarkable event confirming validity of SM, a quantum-trivial local quantum theory of scalar field is, after all, not quantum trivial if the SM is a sector of a non-Abelian gauge theory; analogous to QED event.

Groundbreaking of Supersymmetric Non-Abelian QFTs as well as Integrable QFTs \cite{reshetikhin1981,reshetikhin1983,reshetikhin1985,reshetikhin1987,reshetikhin1987classification,reshetikhin1989,reshetikhin1990,reshetikhin1994,reshetikhin2015} in earnest search for quantum theory of gravity will naturally complement superstring theory. Under such sophistication for superstring theory, which is almost surely strongest pick for fundamental theory of nature, the ultimate truth, it will not be impossible to view all QFTs as effective (low-energy) theory given by renormalizable and nonrenormalizable QFTs, respectively. In other words, every field theory will be a limit in superstring theory. Hypothetically, bosonic strings will be given consideration if the tachyon degrees of freedom form a condensate which is consistently separable by physical expressions; if a perfect factorization, not known to date, is found \cite{mohaupt2003,polchinski1998strings}.

QFT, symmetrical and fundamental, is consistent in framework nevertheless; superstring theory is field theory. A Euclidean nonlocal QFT with nonpolynomial interaction Lagrangian is but robust theory, mathematically rigorous and logically closed \cite{efimov1970nonlocal,efimov1977nonlocal,efimov1985problems,petrina1970ksequations,rebenko1972qed,basuev1973conv,basuev1975convYuk}, dual to statistical physics models, encouraging among other things, further study \cite{brydges1999review,rebenko1988review,brydges1980cmp,polyakov1987gauge,polyakov1977quark,samuel1978grand,o1999duality} by statistical physics analogy which, in the reverse direction, informs every structure of form factors for nonlocality. Under the robust QFT, existence of nonlocality, be it fundamental or a phenomenon, in special case of interest four-dimensional spacetime, is undeniable by all uncertainties. Question of how analytical continuation to the Minkowski spacetime is arranged is, however, open due to lack of existence of the no-go theorem. \cite{efimov1977nonlocal,efimov1985problems}.

At the auspicious moment, really, nonlocal QFT is a self-consistent theory to ensure observed processes can or cannot, on the contrary, be explained. The nonlocal form factor introduced from the physical point of view accounts for meaningful physical processes at too small distances but oversight of experimental design. And, QFT problem is considered solved if mathematical apparatus is created for calculus of the $\mathcal{S}$-matrix of the theory, which is the set of all probability amplitudes of the possible transitions between the states of physical system under consideration.

In hadron interactions (light hadrons) low-energy physics, nonlocal QFT is called nonlocal quark theory, the Virton-Quark model \cite{efimov1985problems,ivanov1981nonlocal,ivanov1989confinement,ivanov1993quark,efimov2001amplitudes,efimov2004blokhintsev} considered effective theory for describing quark confinement field, due to no additional field, typically gluon field, required to ensure quark confinement. While by first-principle QCD, it is impossible to obtain satisfactory description of low-energy hadronic interactions, studies use the robust, original hypothesis that quarks do not exist as arbitrary physical particles, but exist only in virtual state of quasiparticles. The virton field, in framework of QFT, satisfies two conditions: field of free state is identically zero, and causal Green function, the field propagator, is non-zero. That is, non-observability (or non-existence) simply means identically zero-field free-virton; free virton does not exist, and virtons exist only in virtual state. In framework of functional integral, there is nontrivial generating functional for the theory, in particular, for $\mathcal{S}$-matrix of the theory.

Further generalizations of nonlocal QFT, in particular, nonlocal quark theory, include interaction of electromagnetic field with virtons. Moreover, nonlocal QFT arises, also, in functional (nonperturbative, exact) renormalization group (FRG) \cite{kopbarsch,wipf2012statistical,rosten2012fundamentals,igarashi2009realization}. Ultraviolet form factors are functions of differential operators (in coordinate representation) and corresponds to FRG regulators, the regulators of FRG flow of different generating functionals in QFT and statistical physics. While the FRG regulators are not always chosen for entire analytic functions, they are the most preferred for equations of FRG flow with best analytical properties.

Fundamental contributions to formation and development of nonlocal QFT were made, thanks to Gariy Vladimirovich Efimov, in his earlier papers devoted to local QFT with nonpolynomial interaction Lagrangians \cite{efimov1977nonlocal,efimov1963construction,efimov1965nonlinear}, and later papers devoted to nonlocal QFT \cite{efimov1970nonlocal,efimov1967non,efimov1970essentially,efimov1975proof}. His earlier study was developed in parallel but independently in papers of Efim Samoylovich Fradkin \cite{fradkin1963application,fradkin1966application,fradkin2007selected}; hence, the Efimov--Fradkin Theory, which today in its own right is a subject of study by several authors \cite{volkov1968cmp,volkov1969cmp,volkov1970tmp,zumino1969,salam1969,salam1970,fivel1970,graffi1972,horvath1972,wataghin1973,sudarshan1973}. Furthermore, alternatives of the nonlocality were studied by several authors \cite{basuev1974summPT,BelokurovI,BelokurovII,korsun1992VPT,kazakov1979continuation,kazakov1981method,kazakov1988equations}. The idea of nonlocality also was developed in a series of papers by J.W. Moffat and co-authors in the early 90s \cite{moffat1989,moffat1990,ekmw1991,moffat1991,moffat1994}, as well as in the 10s of our century \cite{moffat2011gravity,moffat2011higgs,moffat2016}, in particular, in the context of quantum gravity.

Proof of unitarity of the $\mathcal{S}$-matrix in nonlocal QFT \cite{alebastrov1973proof}, and causality in nonlocal QFT \cite{alebastrov1974proof} were done by G.V. Efimov together with Valeriy Alekseevich Alebastrov. Moreover, in what is considered pinnacle \cite{efimov1977cmp,efimov1979cmp} of set of papers devoted to the nonlocal QFT, G.V. Efimov introduced and investigated the notion of representations of $\mathcal{S}$-matrix of QFT on a discrete lattice of basis functions for both the nonpolynomial \cite{efimov1977cmp} and the polynomial \cite{efimov1979cmp} theory cases. The G.V. Efimov's research is described in three detailed monographs \cite{efimov1977nonlocal,efimov1985problems,ivanov1993quark} brilliantly, not catastrophically, but also in a review \cite{efimov2004blokhintsev}.

A remarkable feature of the chosen G.V. Efimov papers \cite{efimov1977cmp,efimov1979cmp} (see also \cite{efimov1985problems}) is the formulation technique, preserving investigation of strong coupling mode in $\mathcal{S}$-matrix of different QFTs. Moreover, any strong coupling method is as good as gold within QFT, even today. The $\mathcal{S}$-matrix of the nonlocal QFT with nonpolynomial interaction Lagrangian is reducible into grand canonical partition function, the object well-known in statistical physics. This implies a paradigm shift, with the new face of results at crossover of two fundamental sciences.

As in the Efimov's papers, $\mathcal{S}$-matrix studied in this paper in Euclidean metric is applicable to finite volume $V$ (QFT in box) study. For a nonlocal theory, basis functions are eigenfunctions of quantum particle in $D$-dimensional space Schr\"{o}dinger equation, with multidimensional infinite deep well potential terms, on which further expansions are carried out. Limit of infinite volume is taken at end of derivations; and, following necessary renormalization, final expressions for different physical quantities are obtained. The approach is analog of statistical physics model, where corresponding limit is called thermodynamic limit. But, all summations are replaced by integration, for instance, by Poisson summation, in particular, by Euler--Maclaurin formula.

Consciously, drawing attention to the Efimov's papers \cite{efimov1977cmp,efimov1979cmp} (see also \cite{efimov1985problems}) again, we zero in on fundamental differences with this paper: We consider QFT in the whole space initially. At same time, on-and-off switching interaction function $g$ is not constant. That is, coupling constant is an infrared smooth function, chosen such that, for large values of spatial coordinate $x$, the interaction Lagrangian vanishes. This approach is similar to fundamental method described in the monograph of Bogoliubov and Shirkov \cite{bogoliubov1980introduction,bogolyubov1983quantum} and quite nontrivial, since the function $g$ does not change any fundamental property of the phase space system, namely: momentum variable $k$ runs through continuous range of values, and summation over momentum is the initial integral. Moreover, we choose for all expansion basis functions, Hermite basis functions given by eigenfunctions of quantum particle in $D$-dimensional space Schr\"{o}dinger equation with multidimensional isotropic oscillator potential. This choice is also similar to what is described in the monograph by Bogoliubov and others \cite{bogolyubov1990GeneralQFT}. Resulting $\mathcal{S}$-matrix is in representation of basis functions on a discrete lattice of functions.

Another next distinct feature of this paper is we do not adopt any basis at all initially. The theory begins with an $\mathcal{S}$-matrix expression, more precisely, the related generating function $\mathcal{Z}$ in terms of abstract functional integral over the (scalar) primary field \cite{kopbarsch,vasil2004field,zinn1989field}. By formal algebraic operations, the prior is represented as a grand canonical partition function. For the nonpolynomial theory, the resulting partition function is a series over the interaction constant with a finite radius of convergence. And, in principle, it is possible to establish majorizing and minorizing series, the majorant and minorant, as in nonpolynomial Euclidean QFT such as the sine-Gordon model, or in the case of polynomial term modulated by Gaussian function of field \cite{efimov1985problems}. As a result, the $\mathcal{S}$-matrix is studied through statistical physics methods. In the case of the polynomial QFT, resulting partition function is understood in the terms of a formal power series over the interaction constant.

The obtained series, in both the polynomial and nonpolynomial cases, is but intermediate step in derivation of expression for the $\mathcal{S}$-matrix in the representation of basis functions. Further derivation for the desired expression for $\mathcal{S}$-matrix of the theory, in representation of basis functions, is done with the expansion of propagator of the free theory $G$ into an infinite series over a separable basis of Hermitian functions from the $\mathcal{S}$-matrix expression in terms of the partition function. Moreover, such decomposition always exists, since propagator of the free theory $G$ is a square-integrable function, the well-known fact of Hilbert space theory.

Nonlocal QFT is then formulated with basis functions parameterized by continuous variable $k$. Practically repeating derivations, the expression is gotten for $\mathcal{S}$-matrix of the theory: the $\mathcal{S}$-matrix in representation of basis functions on a continuous lattice of functions. The advantage of this approach is that, in all integrals with respect to momentum variable $k$, the integration measure $d\mu$ is treated as arbitrary variable which is intuitively, mathematically correct since $d\mu$ is a formal variable, regardless of its actual value and meaning. The resulting $\mathcal{S}$-matrix, in representation of basis functions on a continuous lattice of functions, is a mathematically defined object for different mathematical analysis techniques, for instance, inequality methods.

Jensen inequality, the central inequality of this paper, is used, in particular, to formulate resulting $\mathcal{S}$-matrix (strictly speaking, $-\ln{\mathcal{S}}$) in terms of variational principle: constructing majorant of corresponding lattice integral in terms of variational function $q$, and minimizing the resulting majorant with respect to $q$; a well-known technique, for instance, in polaron problem in terms of functional integral (polaronics \cite{nedelko1995oscillator}). However, we note that in continuous case, upon applying Jensen inequality, additional divergences arise due to structure of Gaussian integral measure $D\sigma$ (not to be mixed with $d\mu$) on continuous lattice of variables; but, plausibility of a representation on continuous lattice of functions is not negated. 

Moreover, the presented divergence problem, in principle, is solved by the FRG method: Finding the FRG flow regulator $R_{\varLambda}$ containing  measure $d\mu$ as its argument, and carrying out formal derivations such that renormalization (rescaling) of strictly defined ``building blocks of theory'' in terms of the physical equivalents is done in final derivation; an approach which is analogous to the Wilson RG, consisting of mainly two steps, decimation and rescaling \cite{kopbarsch}. But the approach leaves a number of questions unanswered: What is then the connection of original $\mathcal{S}$-matrix of the theory to representation in basis functions on continuous lattice of functions? Is it possible to redefine an original expression for $\mathcal{S}$-matrix by rescaling the same strictly defined building blocks of the theory? And, in what sense can the Jensen inequality be further understood?

The FRG approach reminds us of ``Cheshire Cat'', the mischievous grin in suggestion of how the $\mathcal{S}$-matrix of the theory is to be defined from the very beginning. As a result, we then give an alternative \emph{formulation} for the $\mathcal{S}$-matrix of the theory where results on a discrete or continuous lattice of basis functions are consistent in both polynomial and nonpolynomial QFTs, with no question that the mathematical theory is both rigorous and closed.

In summary, the motivation and methodology of this paper have been described. In the next section, the methodology is implemented in framework of generalized theory. In section three, the focus is on polynomial $\varphi^{4}$ theory in $D$-dimensional spacetime. The variational principle is restricted to the case satisfying separable kernels of the resulting lattice and integral formulae; this approach increases the value of the majorant, which is consistent with the framework. In latter part of the section, the case of polynomial $\varphi^{6}$ theory in $D$-dimensional spacetime is discussed. In section four, nonpolynomial sine-Gordon theory in $D$-dimensional spacetime are studied; separable kernels of the lattice and integral equations are considered. 

In section five, numerical results are presented for prior obtained analytical expressions. To simplify computations, the case $D=1$ is observed, although analytical inferences obtained do not only apply to $D=1$, in principle. Further simplifying computations, the Gaussian function is chosen as propagator for the free theory $G$, since such model is close to reality: the propagator $G$ in the form of Gaussian function, in particular, is typical in the Virton-Quark model, which is the generally accepted model for describing light hadrons low-energy physics \cite{efimov1985problems,ivanov1981nonlocal,ivanov1989confinement,ivanov1993quark,efimov2001amplitudes,efimov2004blokhintsev}. Moreover, with the Gaussian choice of propagator, final expressions only arrive in simple, transparent form. 

In latter part of section five, we give an original proposal, consisting in the main of the reinterpreting the concept of nonlocal QFT. We propose considering the theory in internal space with extra dimensions and subsequent compactification into physical four-dimensional spacetime. This not only changes the meaning of nonlocal interaction ultraviolet length parameter $l$, the needle of pivot for ratio of ultraviolet and infrared parameters in the theory, it also changes analytical properties of the Green functions, scattering amplitudes, and form factors, in terms of physical variables obtained following compactification from internal space with extra dimensions; yet, method compactification is determined by process \cite{mohaupt2003,polchinski1998strings,west2016extracoords}. In other words, in section five, we propose to change the concept of connections in nonlocal QFT. In the conclusion, a lasting discussion on all obtained results and inferences is left.
\section{General Theory}

\subsection{Derivation of the $\mathcal{S}$-Matrix in Terms of the Grand Canonical Partition Function}

In this subsection, we obtain an expression for the $\mathcal{S}$-matrix in terms of the grand canonical partition function, which in turn is to be expressed in terms of \emph{the abstract functional integral} over the primary fields of the theory.

Let the generating functional $\mathcal{Z}$ be given by an abstract functional integral for the free (Gaussian) theory, similar to Gaussian QFT with classical Gaussian action \cite{reshetikhin2015}. That is, $\mathcal{Z}$ is given by \cite{kopbarsch,vasil2004field,zinn1989field,nedelko1995oscillator}:
\begin{equation}
  \mathcal{Z}[j]=\int \mathcal{D}[\varphi] e^{-S_{0}[\varphi] + (j|\varphi)}=\mathcal{Z}_{0} e^{\frac{1}{2}(j|\hat G|j)}
\end{equation}
where $S_{0}[\varphi]=\frac{1}{2}(\varphi|\hat L|\varphi)$ is action of the free theory, $\hat L=\hat G^{-1}$ is inverse propagator, $j=j^{'}+ij^{''}\in\mathbb{C}$ is source, which in general is a complex field, $(j|\varphi)=\int d^{D}z j(z)\varphi(z)$ is scalar product of the source and primary field (the definition is given for real-valued $j$ and $\varphi$ configurations), and the quadratic form of the source is generally given by integration over $x$ and $y$: $(j|\hat G|j)=\int d^{D}x\int d^{D}y j(x)G(x-y)j(y)$.

The generating functional $\mathcal{Z}$ of the theory with interaction is written in the form \cite{kopbarsch,vasil2004field,zinn1989field,nedelko1995oscillator}:
\begin{equation}
    \mathcal{Z}[g,j]=\int \mathcal{D}[\varphi] 
    e^{-S_{0}[\varphi]-S_{1}[g,\varphi] + (j|\varphi)}
    \label{generation_function_interaction}
\end{equation}
where $S_{1}[g,\varphi]$ is interaction action, the action responsible for system interaction, given by \cite{efimov1977nonlocal,efimov1985problems}:
\begin{equation}
   S_{1}[g,\varphi]= \int d^{D}x g(x)U[\varphi(x)] = \int d^{D}x g(x)\int\limits_{-\infty}^{+\infty}\frac{d\lambda}{2\pi}\tilde{U}(\lambda)e^{i\lambda \varphi(x)} \equiv \int d\varGamma e^{i\lambda \varphi(x)}
\end{equation}
such that $\varGamma$ is a notation for a point in the system phase space, which is the product of spaces $\lambda$ and $x$; the class of the function $\tilde{U}(\lambda)$ does not play a role, up to certain point; interaction Lagrangian $\tilde{U}(\lambda)$ is a distribution, in general; and, the Fourier transform of $\tilde{U}(\lambda)$ is not a function, in a classical sense, for instance, $\tilde{U}(\lambda)$ is a polynomial in the primary field $\varphi$.

The generating functional $\mathcal{Z}$ in (\ref{generation_function_interaction}) expands into functional Taylor series over coupling constant $g$:
\begin{equation} \label{expansion}
    \mathcal{Z}[g,j]=\int \mathcal{D}[\varphi]e^{-S_{0}[\varphi] + (j|\varphi)} \sum\limits_{n=0}^{\infty}\frac{(-1)^{n}}{n!}\left\{\prod\limits_{a=1}^{n} \int d\varGamma_{a}\right\}e^{i\sum\limits_{a=1}^{n} \lambda_{a}\varphi(x_{a})}
\end{equation}
Therefore, the following representation is valid:
\begin{equation*}
  i\lambda_{a}\varphi(x_{a}) = \int d^{D}z i\lambda_{a}\varphi(z)\delta^{(D)}(z-x_{a}) =  \lb i\lambda_{a}\delta_{\bullet x_{a}}\big|\varphi \rb
\end{equation*}
where scalar product is defined in terms of integral; and, notation $X_{\bullet}$ means variable $X$ without explicit argument, for instance, the result of integration which is independent of the integrand variable.

By the resulting representation, expression (\ref{expansion}) is given by:
\begin{equation*} 
\begin{split}
 &\mathcal{Z}[g,j]=\sum\limits_{n=0}^{\infty}\frac{(-1)^{n}}{n!} \lbc\prod\limits_{a=1}^{n} \int d\varGamma_{a}\rbc\int \mathcal{D}[\varphi] e^{-S_{0}[\varphi] + \left(j+i\sum\limits_{a=1}^{n}\lambda_{a}\delta_{\bullet x_{a}}\big|\varphi\right)} = \\ &= \sum\limits_{n=0}^{\infty}\frac{(-1)^{n}}{n!} \lbc\prod\limits_{a=1}^{n} \int d\varGamma_{a}\rbc\mathcal{Z}_{0} e^{\frac{1}{2} \lb j+i\sum\limits_{a=1}^{n}\lambda_{a}\delta_{\bullet x_{a}}\big|\hat G\big|j+i\sum\limits_{b=1}^{n}\lambda_{b}\delta_{\bullet x_{b}}\rb} = \\ &= \sum\limits_{n=0}^{\infty}\frac{(-1)^{n}}{n!} \lbc \prod\limits_{a=1}^{n} \int d\varGamma_{a}\rbc\lbc\mathcal{Z}_{0} e^{\frac{1}{2} (j|\hat G|j)}\rbc e^{-\frac{1}{2}\sum\limits_{a,b=1}^{n}\lambda_{a}\lambda_{b}G(x_{a}-x_{b}) + i\sum\limits_{a=1}^{n}\lambda_{a}\hat Gj(x_{a})} = \\&=  \mathcal{Z}[j]\sum\limits_{n=0}^{\infty}\frac{(-1)^{n}}{n!} \lbc \prod\limits_{a=1}^{n} \int d\varGamma_{a}\rbc e^{-\frac{1}{2}\sum\limits_{a,b=1}^{n}\lambda_{a}\lambda_{b}G(x_{a}-x_{b}) + i\sum\limits_{a=1}^{n}\lambda_{a}\hat Gj(x_{a})}.
\end{split}
\end{equation*}

The result of action of the operator $\hat{G}$ on the source $j$ is given by
\begin{equation*}
    \hat G j (x_{a}) \equiv \int d^{D}z G(z-x_{a})j(z) \equiv \bar \varphi(x_{a}).
\end{equation*}

We now introduce a central object of QFT: the $\mathcal{S}$-matrix of the theory, given in terms of functional $\mathcal{Z}$ by an explicit expression (similar to \cite{efimov1977nonlocal,efimov1985problems}):
\begin{equation}\label{s_matrix}
    \mathcal{S}[g,\bar \varphi] \equiv \frac{\mathcal{Z}[g,j=\hat G^{-1}\bar\varphi]}{\mathcal{Z}[j=\hat G^{-1}\bar\varphi]} = \sum\limits_{n=0}^{\infty}\frac{(-1)^{n}}{n!} \lbc\prod\limits_{a=1}^{n} \int d\varGamma_{a}e^{i \lambda_{a}\bar \varphi(x_{a})}\rbc e^{-\frac{1}{2}\sum\limits_{a,b=1}^{n}\lambda_{a}\lambda_{b}G(x_{a}-x_{b})}.
\end{equation}

In particular, by above expression (\ref{s_matrix}), the $\mathcal{S}$-matrix of the theory is given in terms of the grand canonical partition function. If Fourier transform of the interaction Lagrangian $\tilde{U}(\lambda)$ is classical, for instance, quadratically integrable function, the resulting partition function is a series with finite radius of convergence, and majorizing and minorizing series, the majorant and the minorant, respectively, are obtainable, as in nonpolynomial Euclidean QFT, for instance, in sine-Gordon model, or when the term polynomial in the field is modulated by the Gaussian function. Consequently, the $\mathcal{S}$-matrix of the theory is studied by statistical physics method.

If original interaction Lagrangian is polynomial in the primary field $\varphi$, the Fourier transform of $\tilde{U}(\lambda)$ is a distribution; and, expression (\ref{s_matrix}) is a \emph{formal power series with respect to the interaction constant}, on an expansion of the propagator $G$ into an infinite series over a separable basis. In particular, such decomposition always exists, since the propagator $G$ is square integrable function, a fact well known from theory of Hilbert spaces. In the next subsection, we discuss the propagator splitting.

\subsection{Propagator Splitting}

The expansion of the propagator of the free theory $G$ in an infinite series over a separable basis is given by the following steps: First, $G$ is represented in the form (similar to \cite{efimov1985problems,efimov1977cmp,efimov1979cmp})
\begin{equation}
    G(x-y) = \int d^{D}z D(x-z)D(z-y).
\end{equation}
That is, if function $D$ is even, then $\int d^{D}z D^{2}(z) = G(0) < \infty$. This implies $D(z)\in L_{2}\left(\mathbb{R}^{D}\right)$, for all nonlocal QFTs. 

Now, in the second step, we introduce the notation
\begin{equation}
    D_{s}(x) \equiv \int d^{D}z D(x-z)\psi_{s}(z), \quad s=(s_{1},...,s_{D}),\quad  s_{i}=0,1,2,...
\end{equation}
for coefficients of expansion of the function $D(z)$ over the basis of functions $\psi_{s}$
\begin{equation*}
    \psi_{s}(z) = \psi_{s_{1}}(z_{1})\cdot\ldots \cdot\psi_{s_{D}}(z_{D}),
\end{equation*}
where the building blocks of a basis are functions $\psi_{s_{i}}(z_{i})$, the one-dimensional Hermite functions known in theory of a quantum harmonic oscillator, explicitly given by:
\begin{equation}
    \psi_{s_{i}}(z_{i})=\frac{1}{\sqrt{2^{s_{i}}s_{i}!\sqrt{\pi}}}
    H_{s_{i}}(z_{i})e^{-\frac{1}{2}z_{i}^{2}}.
\end{equation}

Recall the completeness condition for basic functions
\begin{equation}
    \sum\limits_{s}\psi_{s}(z)\psi_{s}(z') =\delta^{(D)}(z-z')
\end{equation}
elementarily reduced to one-dimensional analogue, since 
\begin{equation*}
    \sum\limits_{s} \equiv \sum\limits_{s_{1}=0}^{\infty}\ldots
    \sum\limits_{s_{D}=0}^{\infty},\quad
    \sum\limits_{s}^{N} \equiv \sum\limits_{s_{1}=0}^{N}\ldots
    \sum\limits_{s_{D}=0}^{N}
\end{equation*}
where the second equivalence is valid if we truncate the lattice with truncation parameter $N$ (what is used in numerical calculations). Next, consider the chain of equalities:
\begin{equation*}
\begin{split}
    &\sum\limits_{s}D_{s}(x)D_{s}(y) =    \sum\limits_{s} \int d^{D}z D(x-z)\psi_{s}(z)\int d^{D}z' D(y-z')\psi_{s}(z') =\\&= \int d^{D}z\int d^{D}z'
    D(x-z)D(y-z')\sum\limits_{s}\psi_{s}(z)\psi_{s}(z')\\ &=  \int d^{D}z\int d^{D}z' D(x-z)D(y-z')\delta^{(D)}(z-z') \\ &= \int d^{D}z D(x-z)D(z-y) = G(x-y).
\end{split}
\end{equation*}
As a result, an expression is obtained for the propagator $G$ in the form of (infinite) sum of products of functions, each depending only on one variable ($x$ or $y$):
\begin{equation}\label{SepExpForProp}
    \boxed{G(x-y)=\sum\limits_{s}D_{s}(x)D_{s}(y)}\, .
\end{equation}
The expression (\ref{SepExpForProp}) is the desired decomposition of propagator of the free theory $G$ into an infinite series over a separable basis. The expression will now be used in the next subsection.

\subsection{Derivation of the $\mathcal{S}$-Matrix in the Representation of Basis Functions (Discrete Lattice of Functions)}

In this subsection, we carry out a rather different derivation for the $\mathcal{S}$-matrix of the theory: From representation for $\mathcal{S}$ in terms of the grand canonical partition function we derive representation for $\mathcal{S}$ in terms of the basis functions. The latter is well defined in the case where Fourier transform of $\tilde{U}(\lambda)$ is a distribution, which is the case for polynomial field theories. 

To obtain desired representation, we apply the method of propagator splitting to expression (\ref{s_matrix}):
\begin{equation}
\begin{split}
   &e^{-\frac{1}{2}\sum\limits_{a,b=1}^{n}\lambda_{a}\lambda_{b}G(x_{a}-x_{b})} 
   = e^{-\frac{1}{2}\sum\limits_{a,b=1}^{n}\lambda_{a}\lambda_{b}\sum\limits_{s}D_{s}(x_{a})D_{s}(x_{b})} 
   = e^{\sum\limits_{s}\lbc-\frac{1}{2}\sum\limits_{a=1}^{n}\sum\limits_{b=1}^{n}\lambda_{a}D_{s}(x_{a})\lambda_{b}D_{s}(x_{b})\rbc} = 
   \\[.1in] 
   &= \prod\limits_{s}e^{-\frac{1}{2}\lbc\sum\limits_{a=1}^{n}\lambda_{a}D_{s}(x_{a})\rbc\lbc\sum\limits_{b=1}^{n}\lambda_{b}D_{s}(x_{b})\rbc} 
   \equiv \prod\limits_{s}e^{-\frac{1}{2}\xi_{s}\xi_{s}} 
   = \prod\limits_{s}e^{\frac{1}{2}(i\xi_{s})^{2}} 
   = \prod\limits_{s}\int\limits_{-\infty}^{+\infty}\frac{dt_{s}}{\sqrt{2\pi}}e^{-\frac{1}{2}t^{2}_{s}+i\xi_{s}t_{s}}
\end{split}
\end{equation}
where the product over $s$ is in the standard multiplicative form 
\begin{equation*}
    \prod\limits_{s} \equiv \prod\limits_{s_{1}=0}^{\infty}\ldots
    \prod\limits_{s_{D}=0}^{\infty},\quad\quad
    \prod\limits_{s}^{N} \equiv \prod\limits_{s_{1}=0}^{N}\ldots
    \prod\limits_{s_{D}=0}^{N}.
\end{equation*}
The expression (\ref{s_matrix}) for the $\mathcal{S}$-matrix of the theory then takes the form:
\begin{equation}
    \mathcal{S}[g,\bar \varphi] = \sum\limits_{n=0}^{\infty}\frac{(-1)^{n}}{n!} \lbc\prod\limits_{a=1}^{n} \int d\varGamma_{a}e^{i \lambda_{a}\bar \varphi(x_{a})}\rbc\prod\limits_{s}\int\limits_{-\infty}^{+\infty}\frac{dt_{s}}{\sqrt{2\pi}}e^{-\frac{1}{2}t^{2}_{s}+it_{s}\sum\limits_{a=1}^{n}\lambda_{a}D_{s}(x_{a})}.
\end{equation}
Furthermore, the following chain of long but simple equalities holds:
\begin{equation}
\begin{split}
    \mathcal{S}[g,\bar \varphi] &= \sum\limits_{n=0}^{\infty}\frac{(-1)^{n}}{n!} \lbc\prod\limits_{a=1}^{n} \int d\varGamma_{a}e^{i \lambda_{a}\bar \varphi(x_{a})}\rbc\prod\limits_{s}\int\limits_{-\infty}^{+\infty}\frac{dt_{s}}{\sqrt{2\pi}}e^{-\frac{1}{2}t^{2}_{s}+it_{s}\sum\limits_{a=1}^{n}\lambda_{a}D_{s}(x_{a})} = \\ &= \sum\limits_{n=0}^{\infty}\frac{(-1)^{n}}{n!} \lbc\prod\limits_{s}\int\limits_{-\infty}^{+\infty}\frac{dt_{s}}{\sqrt{2\pi}}e^{-\frac{1}{2}t^{2}_{s}} \rbc \lbc\prod\limits_{a=1}^{n} \int d\varGamma_{a}\rbc  e^{i\sum\limits_{a=1}^{n}\lambda_{a}\left[\bar \varphi(x_{a})+\sum\limits_{s}t_{s}D_{s}(x_{a})\right]} =\\
    &=\sum\limits_{n=0}^{\infty}\frac{(-1)^{n}}{n!} \lbc\prod\limits_{s}\int\limits_{-\infty}^{+\infty}\frac{dt_{s}}{\sqrt{2\pi}}e^{-\frac{1}{2}t^{2}_{s}} \rbc \lbc\prod\limits_{a=1}^{n} \int d\varGamma_{a} e^{i\lambda_{a}\left[\bar \varphi(x_{a})+\sum\limits_{s}t_{s}D_{s}(x_{a})\right]}\rbc = \\
    &= \sum\limits_{n=0}^{\infty}\frac{(-1)^{n}}{n!} \lbc\prod\limits_{s}\int\limits_{-\infty}^{+\infty}\frac{dt_{s}}{\sqrt{2\pi}}e^{-\frac{1}{2}t^{2}_{s}} \rbc \lbc S_{1}\left[\bar \varphi_{\bullet}+\sum\limits_{s}t_{s}D_{s,\bullet}\right]\rbc^{n} =\\
    &=  \lbc \prod\limits_{s}\int\limits_{-\infty}^{+\infty}\frac{dt_{s}}{\sqrt{2\pi}}e^{-\frac{1}{2}t^{2}_{s}} \rbc  e^{-S_{1}\left[\bar \varphi_{\bullet}+\sum\limits_{s}t_{s}D_{s,\bullet}\right]}.
    \end{split}
\end{equation}
The chain is based on transformation of each replica (a term with fixed variable $s$) in terms of the usual Gaussian integral, where each integral provides a new lattice integration variable $t_{s}$, the number of which is infinite, such that lattice is space of functions. Moreover, compared to original abstract functional integral, this infinite integral has several advantages, for instance, providing the easy to prove existence and uniqueness of a given integral in  mathematically rigorous sense. Therefore, the latter can be investigated with mathematical techniques, for instance, the methods of inequalities (in particular, theorem of two policemen).

By the transformation chain, the $\mathcal{S}$-matrix of the theory is given by the expression (as in \cite{efimov1985problems,efimov1977cmp,efimov1979cmp}):
\begin{equation}
\boxed{ \mathcal{S}[g,\bar \varphi] =\lbc\prod\limits_{s}\int\limits_{-\infty}^{+\infty}
\frac{dt_{s}}
{\sqrt{2\pi}}e^{-\frac{1}{2}t_{s}^{2}}\rbc e^{-\int d^{D}xg(x)
U\left[\bar \varphi(x) + \sum\limits_{s}t_{s}D_{s}(x)\right]}}\, .
\label{Smatrix_basis_function_representation}
\end{equation}
That is the $\mathcal{S}$-matrix is obtained in the \emph{representation of basis functions}. And, now, remarkably, the expression (\ref{Smatrix_basis_function_representation}) for the $\mathcal{S}$-matrix is represented as a limit of sequence $\{\mathcal{S}_{N}\}_{0}^{\infty}$ of functionals:
\begin{equation*}
\mathcal{S}[g,\bar \varphi]=\lim\limits_{N\rightarrow\infty}
\mathcal{S}_{N}[g,\bar \varphi] 
\end{equation*}
with common member given by 
\begin{equation}
\mathcal{S}_{N}[g,\bar \varphi] =\lbc \prod\limits_{s}^{N}\int\limits_{-\infty}^{+\infty}\frac{dt_{s}}
{\sqrt{2\pi}}e^{-\frac{1}{2}t_{s}^{2}}\rbc e^{-\int d^{D}xg(x)
U\left[\bar \varphi(x) + \sum\limits_{s}t_{s}D_{s}(x)\right]}.
\end{equation}
In general, the prelimit expression for the $\mathcal{S}$-matrix is convenient both for theoretical study and numerical calculations. In the next subsection, we consider the variational principle for the resulting $\mathcal{S}$-matrix: Namely, we construct a majorant of corresponding lattice integral, depending on function $q_{s}$ of discrete variable $s$; and, minimize the majorant with respect to function $q_{s}$. The technique is well known, for instance, in polaron problem, in terms of the functional integral (polaronics).

\subsection{Majorant and Variational Principle}

The variational principle begins with identity transformation on expression (\ref{Smatrix_basis_function_representation}):
\begin{equation}
     \mathcal{S}[g,\bar \varphi] =\lbc \prod\limits_{s}\int\limits_{-\infty}^{+\infty}\frac{dt_{s}}
{\sqrt{2\pi}}e^{-\frac{1}{2}(1+q_{s})t_{s}^{2} + \frac{1}{2} q_{s}t_{s}^{2}}\rbc e^{-\int d^{D}xg(x)
U\left[\bar \varphi(x) + \sum\limits_{s}t_{s}D_{s}(x)\right]}.
\end{equation}
Using change of variables:
\begin{equation*}
    t_{s} = \frac{u_{s}}{\sqrt{1+q_{s}}}, \quad dt_{s} = \frac{du_{s}}{\sqrt{1+q_{s}}} 
\end{equation*}
further identical transformations of the expression for the $\mathcal{S}$-matrix are carried out as follows:
\begin{equation}
\begin{split}
     \mathcal{S}[g,\bar \varphi] &=\lbc \prod\limits_{s}\int\limits_{-\infty}^{+\infty}\frac{du_{s}}
{\sqrt{2\pi}}\frac{1}{\sqrt{1+q_{s}}}e^{-\frac{1}{2}u^{2}+\frac{1}{2}\frac{q_{s}}{1+q_{s}}u_{s}^{2}}\rbc e^{-\int d^{D}xg(x)
U\left[\bar \varphi(x) + \sum\limits_{s}u_{s}\frac{D_{s}(x)}{\sqrt{1+q_{s}}}\right]} = \\
& =  \int d\sigma_{u}e^{-\int d^{D}xg(x)
U\lbq\bar \varphi(x) + \sum\limits_{s}u_{s}\frac{D_{s}(x)}{\sqrt{1+q_{s}}}\rbq - \frac{1}{2}\sum\limits_{s}\ln{(1+q_{s})} + \frac{1}{2}\sum\limits_{s}\frac{q_{s}}{1+q_{s}}u_{s}^{2}}
\label{before_einsen}
\end{split}
\end{equation}
where, in second line of expression (\ref{before_einsen}), there is introduction of integral measure 
\begin{equation}
\int d\sigma_{u} \equiv \prod\limits_{s}\int\limits_{-\infty}^{+\infty}\frac{du_{s}}
{\sqrt{2\pi}}e^{-\frac{1}{2}u_{s}^{2}}
\label{EfimovMera}
\end{equation}
called the Efimov measure.

Naturally, all the transformations are identical. We now use the inequality, Jensen inequality, to get the majorant in this method. Notably, while the Jensen inequality is a demonstration of the fact that the exponential function in expression (\ref{before_einsen}) is convex, a good enough variational principle is obtained if appropriate measure of integration is chosen. In particular, with the Efimov measure (\ref{EfimovMera}) in terms of variables $u_{s}$ as the measure of choice, the Jensen inequality is given as follows:
\begin{equation}
     \mathcal{S}[g,\bar \varphi] \geq  e^{-\int d\sigma_{u} \left\{\int d^{D}x g(x)U\lbq\bar \varphi(x)+\sum\limits_{s}u_{s}\frac{D_{s}(x)}{\sqrt{1+q_{s}}}\rbq + \frac{1}{2}\sum\limits_{s}\ln{(1+q_{s})} - \frac{1}{2}\sum\limits_{s}\frac{q_{s}}{1+q_{s}}u_{s}^{2}\right\} }.
\end{equation}
However, strictly speaking, this inequality for the $\mathcal{S}$-matrix gives the minorant. We therefore consider a different functional $\mathcal{G}$, which is (minus) logarithm of the $\mathcal{S}$-matrix. Such functional $\mathcal{G}$ of the field $\bar\varphi$ is the generating functional of the amputated connected Green functions, which are coefficients of the expansion of $\mathcal{G}$ into functional Taylor series over the field $\bar\varphi$. By definition, $\mathcal{G}$ is given by
\begin{equation}
    \mathcal{G}[g,\bar \varphi] = - \ln{\mathcal{S}[g,\bar \varphi]}.
\label{mathcal_G}
\end{equation}
The majorant of $\mathcal{G}$ is given by the expression
\begin{equation}
     \mathcal{G}[g,\bar \varphi] \leq \int d\sigma_{u} \left\{\int d^{D}x g(x)U\left[\bar \varphi(x)+\sum\limits_{s}\frac{u_{s}D_{s}(x)}
     {\sqrt{1+q_{s}}}\right] + \frac{1}{2}\sum\limits_{s}\ln{(1+q_{s})} - \frac{1}{2}\sum\limits_{s}\frac{u_{s}^{2}q_{s}}{1+q_{s}}\right\}.
     \label{G}
\end{equation}
Transforming the terms on the right side of expression (\ref{G}), consider the first term:
\begin{equation}
\begin{split}
    &\int d\sigma_{u} U\left[\bar\varphi(x)+\sum\limits_{s}u_{s}\frac{D_{s}(x)}{\sqrt{1+q_{s}}}\right] = \int d\sigma_{u}
    \int\limits_{-\infty}^{+\infty} \frac{d\lambda}{2\pi}\tilde{U}(\lambda) e^{i\lambda\left[\bar \varphi(x)+\sum\limits_{s}u_{s}\frac{D_{s}(x)}{\sqrt{1+q_{s}}}\right]} = \\ &= \int\limits_{-\infty}^{+\infty} \frac{d\lambda}{2\pi}\tilde{U}(\lambda)e^{i\lambda\bar \varphi(x)} \left\{\prod\limits_{s}\int\limits_{-\infty}^{+\infty}\frac{du_{s}}
{\sqrt{2\pi}}e^{-\frac{1}{2}u_{s}^{2}}\right\} e^{i\lambda\sum\limits_{s}u_{s}\frac{D_{s}(x)}{\sqrt{1+q_{s}}}} = \\ &=\int\limits_{-\infty}^{+\infty} \frac{d\lambda}{2\pi}\tilde{U}(\lambda)e^{i\lambda\bar \varphi(x)} e^{-\frac{1}{2}\lambda^{2}\sum\limits_{s}\frac{D_{s}^{2}(x)}{1+q_{s}}}
\end{split}
\end{equation}
where in the remaining terms, integration with respect to Efimov measure is trivial and goes to unity. The expression (\ref{G}) then takes the form (as in \cite{efimov1985problems,efimov1977cmp,efimov1979cmp}):
\begin{equation}
\boxed{
     \mathcal{G}[g,\bar \varphi] \leq \int d^{D}x g(x)\int\limits_{-\infty}^{+\infty} \frac{d\lambda}{2\pi}\tilde{U}(\lambda)e^{i\lambda\bar \varphi(x)} e^{-\frac{1}{2}\lambda^{2}\sum\limits_{s}\frac{D_{s}^{2}(x)}{1+q_{s}}} + \frac{1}{2}\sum\limits_{s}\left[\ln{(1+q_{s})} - \frac{q_{s}}{1+q_{s}} \right]}\, .
     \label{majoranta}
\end{equation}
Therefore, for the generating functional $\mathcal{G}$, the inequality, called majorant, is obtained. The majorant depends on function $q_{s}$, with respect to which the majorant is minimized. And, further derivations are also possible, but only after specifying the theory, that is, the explicit form of the interaction Lagrangian. However, prior to dwelling on a particular theory within the framework of the developed formalism, consider another, yet, interesting question: the derivation of the $\mathcal{S}$-matrix in the representation of basis functions on a continuous lattice of functions. The next subsection is devoted to this, in particular.

\subsection{Derivation of the $\mathcal{S}$-Matrix in the Representation of Basis Functions (Continuous Lattice of Functions)}

Consider the propagator of the free theory $G$ in the $\mathcal{S}$-matrix expression (\ref{s_matrix}), in the momentum $K$ representation. That is, write in the Fourier integral form \cite{efimov1970nonlocal,basuev1973conv}:
\begin{equation}
    G(x_{a}-x_{b}) = \int \frac{d^{D}k}{(2\pi)^{D}}\tilde{G}(k)e^{ik(x_{a}-x_{b})},\quad \tilde{G}(k)=\frac{F(l^{2}k^{2})}{k^{2}+m^{2}}.
    \label{PropFexp}
\end{equation}
The expression (\ref{PropFexp}) then presents a typical propagator for nonlocal QFT: the function $F$ is the ultraviolet form factor depending on (ultraviolet) length parameter $l$, and $m$ is the infrared mass. Notably, $\tilde{G}$ here is an even function of $k$: $\tilde{G}(-k)= \tilde{G}(k)$. Though the given instance may not be used, consider $\tilde{G}$ to be even in $k$. In this case, write $G$ by Fourier integral over cosines:
\begin{equation}
    G(x_{a}-x_{b}) = \int \frac{d^{D}k}{(2\pi)^{D}}\tilde{G}(k)\cos{\left[k(x_{a}-x_{b})\right]}.
\end{equation}
Expanding the cosine of the difference gives the expression for $G$ in form of integral over the separable basis of sines and cosines. This is the analogue of an expansion in the series of separable basis considered in previous subsections. Notably, however, the trigonometric functions here are not square integrable. But a kind of ultraviolet form factor can always be used if the formalism of the theory is to be heeded.

Next, consider the $\lambda$ quadratic form in the exponent of expression (\ref{s_matrix}). For a chain of equalities using the Fourier integral for $G$, write:
\begin{equation}
\begin{split}
&-\frac{1}{2}\sum\limits_{a,b=1}^{n}\lambda_{a}\lambda_{b} \int \frac{d^{D}k}{(2\pi)^{D}}\tilde{G}(k)\cos{\left[k(x_{a}-x_{b})\right]}=\\
&-\frac{1}{2}\sum\limits_{a,b=1}^{n}\lambda_{a}\lambda_{b} \int \frac{d^{D}k}{(2\pi)^{D}}\tilde{G}(k)\left[\cos{(kx_{a})}\cos{(kx_{b})}+ \sin{(kx_{a})}\sin{(kx_{b})}\right]=\\
&-\frac{1}{2}\int \frac{d^{D}k}{(2\pi)^{D}}\tilde{G}(k)\left\{ \lb\sum\limits_{a=1}^{n}\lambda_{a}\cos{(kx_{a})}\rb^{2} + \lb\sum\limits_{a=1}^{n}\lambda_{a}\sin{(kx_{a})}\rb^{2} \right\}.
\end{split}
\end{equation}
Introduce compact notations:
\begin{equation}
\sum\limits_{a=1}^{n}\lambda_{a}\cos{(kx_{a})} \equiv \frac{\xi(k)}{\sqrt{G(0)}},\quad
\sum\limits_{a=1}^{n}\lambda_{a}\sin{(kx_{a})} \equiv \frac{\eta(k)}{\sqrt{G(0)}}.
\end{equation}
Notably, the $\xi$ and $\eta$ variables are without specific dimension, since the dimension of $\lambda^{2}$ has to compensate for dimension of $G$. The quadratic form is then given by the expression:
\begin{equation}
    -\frac{1}{2}\int \frac{d^{D}k}{(2\pi)^{D}}\frac{\tilde{G}(k)}{G(0)}\lbc \xi^{2}(k) + \eta^{2}(k) \rbc.
\end{equation}
Since Fourier transform of $\tilde{G}(k)$ is positive function, introduce a measure in $k$ space according to the following rule:
\begin{equation}
    \int \frac{d^{D}k}{(2\pi)^{D}}\frac{\tilde{G}(k)}{G(0)} \equiv \int d\mu(k).
\end{equation}
Putting the notations together then rewrite the quadratic form in the exponent of expression (\ref{s_matrix}), and consequently, the exponent itself in compact vivid form. A remarkable property of the resulting expression is that it allows the formulation of so-called \emph{multigral} (product integral):
\begin{equation}
    e^{-\frac{1}{2}\int d\mu(k)\left\{ \xi^{2}(k) + \eta^{2}(k) \right\}} = \prod\limits_{k}e^{-\frac{1}{2} d\mu(k)\left\{ \xi^{2}(k) + \eta^{2}(k) \right\}}.
    \label{Multigral}
\end{equation}
The multigral (\ref{Multigral}) plays a central role in further derivation of the $\mathcal{S}$-matrix in the representation of basis functions on a continuous lattice. But, prior to further advancing in this direction, it will be instructive to draw parallels with the G.V. Efimov papers \cite{efimov1977cmp,efimov1979cmp} alongside statistical physics.

In statistical physics, the assumption is that systems have macroscopic volume $V$. In such a box, different sums for a particular physical quantity are dealt with. And, as limit of $V\rightarrow\infty$ is taken, summation goes into semiclassical integration, and integral expressions arise. This approach was preserved by G.V. Efimov, formulating the $\mathcal{S}$-matrix of the nonlocal QFT in terms of functional integral. In this paper, no initial box $V$ is assumed; however, there is the function which turns interaction on and off. Such a function affects the integral describing the interaction action, but does not affect the structure of the phase space of the theory (continuity of the momentum variable, et cetera).

Now, returning to derivation of the $\mathcal{S}$-matrix in representation of basis functions on a continuous lattice, the exponent with $\xi(k)$ is represented as follows (notably, in the second equality, integration variable $t_{k}$ is changed):
\begin{equation}
    e^{-\frac{1}{2}d\mu(k)\xi^{2}(k)} = \int\limits_{-\infty}^{+\infty}\frac{dt_{k}}{\sqrt{2\pi d\mu(k)}}e^{-\frac{1}{2d\mu(k)}t_{k}^{2} + i \xi(k)t_{k}} = \int\limits_{-\infty}^{+\infty}dt_{k}\sqrt{\frac{d\mu(k)}{2\pi}} e^{-\frac{d\mu(k)}{2}t_{k}^{2} + i \xi(k)d\mu(k)t_{k}}.
\end{equation}
Similar equality holds for exponent with $\eta(k)$. They both are based on representation of corresponding exponentials in terms of arbitrary Gaussian integrals, as in the case of discrete lattice of basis functions. However, $d\mu(k)$ is treated as arbitrary variable. The transformations are not only mathematically correct, but preserves the arbitrariness of the variable $d\mu(k)$, while actual value and meaning does not matter in the moment.

As a result of the transformations, the exponent in expression (\ref{s_matrix}) takes the form:
\begin{equation}
    \prod\limits_{k}\iint\limits_{\mathrm{R}^{2}}dt_{c,k}dt_{s,k}\frac{d\mu(k)}{2\pi}e^{-\frac{1}{2}
    d\mu(k)\lb t^{2}_{c,k}+t^{2}_{s,k} \rb} e^{id\mu(k)\left[\xi(k)t_{c,k} +\eta(k)t_{s,k}\right]}.
\end{equation}
By prior notations, the expression is now given by two families of integration variables, $t_{c,k}$ and $t_{s,k}$:
\begin{equation}
\begin{split}
    \prod\limits_{k}\iint\limits_{\mathrm{R}^{2}}dt_{c,k}dt_{s,k}\frac{d\mu(k)}{2\pi}e^{-\frac{1}{2}
    d\mu(k)\lb t^{2}_{c,k}+t^{2}_{s,k}\rb} e^{id\mu(k)\sqrt{G(0)}\sum\limits_{a=1}^{n}\lambda_{a}\left[\cos{(kx_{a})} t_{c,k} +\sin{(kx_{a})}t_{s,k}\right]} =\\
    \left\{\prod\limits_{k}\iint\limits_{\mathrm{R}^{2}}
    dt_{c,k}dt_{s,k}\frac{d\mu(k)}{2\pi} e^{-\frac{1}{2}d\mu(k)(t^{2}_{c,k}+t^{2}_{s,k})}\right\}e^{i\int d\mu(k)\sqrt{G(0)}\sum\limits_{a=1}^{n}\lambda_{a}\left[\cos{(kx_{a})} t_{c,k} +\sin{(kx_{a})}t_{s,k}\right]}.
    \label{ExpForSMatrix}
\end{split}
\end{equation}
In particular, the second line of this expression (\ref{ExpForSMatrix}) is identified with Gaussian integral measure on a continuous lattice of the variables $t_{c,k}$ and $t_{s,k}$, explicitly. In terms of the integral over the variables $t_{c,k}$ and $t_{s,k}$, the expression for the $\mathcal{S}$-matrix in representation of basis functions is given as follows:
\begin{equation}
     \boxed{ \mathcal{S}[g,\bar\varphi]= \left\{\prod\limits_{k,\sigma}\int\limits_{-\infty}^{+\infty}dt_{\sigma,k}\sqrt{\frac{d\mu(k)}{2\pi}} e^{-\frac{1}{2}d\mu(k)t^{2}_{\sigma,k}}\right\}e^{-\int d^{{\scriptsize D}}x g(x) U\left[\bar \varphi(x) + \sum\limits_{\sigma}\int d\mu(k) \psi_{\sigma}(kx)t_{\sigma,k}\right]}}\, .
     \label{Smatrix_basis_function_repr_contcase}
\end{equation}
The compact notations $\sigma\in\{c,s\}$, $\psi_{s}(kx) = \sin{(kx)}$, and $\psi_{c}(kx) = \cos{(kx)}$ are introduced in the last equality. Therefore, the $\mathcal{S}$-matrix in the representation of basis functions on a continuous lattice is obtained. This expression (\ref{Smatrix_basis_function_repr_contcase}) is the continuous analog of the discrete expression (\ref{Smatrix_basis_function_representation}).

Now, assuming for Gaussian measure of integration, there is the notation
\begin{equation}
    \lbc\prod\limits_{k,\sigma}\int\limits_{-\infty}^{+\infty}
    dt_{\sigma,k}\sqrt{\frac{d\mu(k)}{2\pi}} e^{-\frac{1}{2}d\mu(k)t^{2}_{\sigma,k}}\rbc \equiv \int \mathcal{D}\sigma[t].
\end{equation}
Rewrite the expression (\ref{Smatrix_basis_function_repr_contcase}) for the $\mathcal{S}$-matrix in a more compact vivid form as 
\begin{equation}
    \mathcal{S}[g,\bar\varphi] = \int \mathcal{D}\sigma[t]e^{-\int d^{{\scriptsize D}}x g(x) U\lbq\bar \varphi(x) + \sum\limits_{\sigma}\int d\mu(k)\psi_{\sigma}(kx)t_{\sigma,k}\rbq}.
    \label{Smatrix_contcase_compact}
\end{equation}
In this form, the $\mathcal{S}$-matrix expression is most similar to expression for generating functional $\mathcal{Z}$ in terms of abstract functional integral. However, all quantities appearing in this expression (\ref{Smatrix_contcase_compact}) have rigorous  mathematical definitions formulated in this subsection. As a result, various mathematical techniques can be used, for instance, various methods of inequalities, to analyze the $\mathcal{S}$-matrix in the form (\ref{Smatrix_contcase_compact}).

In the next (final) subsection, we consider variational principle for the resulting $\mathcal{S}$-matrix. Namely, we first construct majorant of the integral (\ref{Smatrix_contcase_compact}), depending on a pair of functions $q_{\sigma,k}$ for the continuous variable $k$. We then minimize the resulting majorant with respect to the functions $q_{\sigma,k}$. In particular, significant difference between continuous case and discrete analogue is noted as follows: naive definition of the $\mathcal{S}$-matrix (\ref{Smatrix_contcase_compact}) still contains the pitfall arising as a result of internal structure of the Gaussian integral measure on continuous lattice of $t_{\sigma,k}$.

\subsection{Redefinition of the $\mathcal{S}$-Matrix and Variational Principle}

To better understand why it is necessary to redefine the $\mathcal{S}$-matrix of the theory (\ref{Smatrix_contcase_compact}), consider linear transformation of the integration variables $t_{\sigma,k}$ in the form $t_{\sigma,k}=A_{\sigma,k}u_{\sigma,k}$. With such a change of variables, the Gaussian measure of integration is transformed as follows:
\begin{equation}
\begin{split}
     \int \mathcal{D}\sigma[t] =
     & \int \mathcal{D}\sigma[u] e^{-\sum\limits_{\sigma}\int\lbq \frac{1}{2}d\mu(k) (A_{\sigma,k}^{2}-1)u_{\sigma,k}^{2}-\ln{(A_{\sigma,k})}\rbq}.
\end{split}    
\end{equation}
Following change of variables, the expression for the $\mathcal{S}$-matrix is given by 
\begin{equation}
     \mathcal{S}[g,\bar \varphi] = \int \mathcal{D}\sigma[u]e^{-S_{1}-\sum\limits_{\sigma}\int\lbq \frac{1}{2}d\mu(k) (A_{\sigma,k}^{2}-1)u_{\sigma,k}^{2}-\ln{(A_{\sigma,k})}\rbq }.
\label{S_matrix_after_linear_change}
\end{equation}
Considering generating functional of amputated connected Green functions $\mathcal{G}$ (\ref{mathcal_G}) and applying Jensen inequality to expression (\ref{S_matrix_after_linear_change}), the majorant of the functional (\ref{mathcal_G}) is given in form of sum of two terms:
\begin{equation}
    \mathcal{G}\lbq g, \bar \varphi\rbq \leq \sum\limits_{\sigma}\int \lb\frac{1}{2}\lb A_{\sigma,k}^{2}-1\rb  - \ln{A_{\sigma,k}} \rb + \int \mathcal{D}\sigma\lbq u\rbq S_{1}.
\label{Cat1}
\end{equation}

By closer examination, the first term on the right-hand side of (\ref{Cat1}) is the $d\mu(k)$-less expression. Such expression can be given in different forms, and, depending on the form, different formulations for the $\mathcal{S}$-matrix of the theory, can be obtained. For instance, considering a version of the RG flow for family of functionals $\mathcal{F}_{\varLambda}$ in RG time $\varLambda$ \cite{kopbarsch,wipf2012statistical,rosten2012fundamentals}:
\begin{equation}
    \mathcal{F}_{\varLambda}\lbq g,\bar\varphi\rbq \equiv \sum\limits_{\sigma}\int d\mu(k) R_{\varLambda}\lbq d\mu(k)\rbq \lb \frac{1}{2}\lb A_{\sigma,k}^{2}-1\rb  - \ln{A_{\sigma,k}} \rb + \int \mathcal{D}\sigma\lbq u\rbq S_{1}.
\label{F_Lambda}
\end{equation}
In the expression (\ref{F_Lambda}), the function (regulator) $R_{\varLambda}$ defines the RG flow; the argument is the measure $d\mu$. If the RG time is zero i.e. $\varLambda=0$, the expression (\ref{F_Lambda}) equals the majorant obtained earlier. The method then gives mathematically correct results both for pair of functions $q_{\sigma,k}$, and for the functional $\mathcal{F}_{\varLambda}$ itself. In other words, this approach is suitable to (formal) derivations, including, in final results, the rescaling of strictly defined building blocks of the theory in terms of physical equivalents, for instance, in terms of the physical measure $d\mu_{ph}=d\mu R_{\varLambda}$. This approach is analogized in the Wilson RG \cite{kopbarsch,wipf2012statistical,rosten2012fundamentals} which consists of two steps: decimation and rescaling.

However, the following question remains important: what is the status of the original $\mathcal{S}$-matrix of the theory in the representation of basis functions (on the continuous lattice of functions). And, is it possible to redefine the original expression for the $\mathcal{S}$-matrix by the rescaling of same strictly defined building blocks of the theory. That is, the approach brings home the ``Cheshire Cat,'' the mischievous grin for implausibility of a redefinition of the $\mathcal{S}$-matrix of the theory right at the start.

Take the \emph{definition} of the $\mathcal{S}$-matrix as follows:
\begin{equation}
     \boxed{ \mathcal{S}[g,\bar\varphi]= \left\{\prod\limits_{k,\sigma}^{d\mu(k)}\int\limits_{-\infty}^{+\infty}\frac{dt_{\sigma,k}}{\sqrt{2\pi}} e^{-\frac{1}{2}t^{2}_{\sigma,k}}\right\}e^{-\int d^{{\scriptsize D}}x g(x) U\left[\bar \varphi(x)+\sum\limits_{\sigma}\int d\mu(k) \psi_{\sigma}(kx)t_{\sigma,k}\right]}}\, .
     \label{Smatrix_bfr_Cat}
\end{equation}
In the expression (\ref{Smatrix_bfr_Cat}), the integration variables $\sqrt{d\mu(k)}t_{\sigma,k}\longrightarrow t_{\sigma,k}$ have been changed; the basic functions $\psi_{\sigma}(kx)/\sqrt{d\mu(k)}\rightarrow\psi_{\sigma}(kx)$ are also redefined, corresponding to the trigonometric functions $\cos{(kx)}$ and $\sin{(kx)}$ from now on. An important point here, however, is that the product over the modes $(k,\sigma)$ is now understood as follows (where $f$ is an arbitrary function):
\begin{eqnarray}
\prod\limits_{k,\sigma}^{d\mu(k)}e^{f_{\sigma}\left(k\right)}=
e^{\sum\limits_{\sigma}\int d\mu\left(k\right)
f_{\sigma}\left(k\right)}.
\label{Cat2}
\end{eqnarray}
From now on, the expression (\ref{Smatrix_bfr_Cat}) is the definition of the $\mathcal{S}$-matrix of a theory on continuous lattice of functions. Variational principle will be formulated on the $\mathcal{S}$-matrix in this form (\ref{Smatrix_bfr_Cat}). The Gaussian measure of integration now has the form
\begin{equation}
    \lbc\prod\limits_{k,\sigma}^{d\mu(k)}\int\limits_{-\infty}^{+\infty}\frac{dt_{\sigma,k}}{\sqrt{2\pi}} e^{-\frac{1}{2}t^{2}_{\sigma,k}}\rbc \equiv \int \mathcal{D}\sigma[t].
\label{Cat3}
\end{equation}
Moreover, the measure (\ref{Cat3}) is also normalized to one. To show the mathematical correctness of the formulated definition of the $\mathcal{S}$-matrix (\ref{Smatrix_bfr_Cat}), results on the discrete and continuous lattices of basis functions, for both polynomial and nonpolynomial QFTs, will be compared.

Returning to the derivation of the majorant, expression (\ref{Cat1}) now has the form
\begin{equation}
    \mathcal{G}\lbq g, \bar \varphi\rbq \leq \sum\limits_{\sigma}\int d\mu(k) \lb\frac{1}{2}\lb A_{\sigma,k}^{2}-1\rb  - \ln{A_{\sigma,k}} \rb + \int \mathcal{D}\sigma\lbq u\rbq S_{1}.
\label{Cat4}
\end{equation}
The required majorant, denoted $\mathcal{F}$, is given in the form of sum of two terms:
\begin{equation}
    \mathcal{F}\lbq g, \bar\varphi\rbq = I_{1} + I_{2},\quad
    I_{1} = \int \mathcal{D}\sigma\lbq u\rbq S_{1}, \quad I_{2} = \sum\limits_{\sigma}\int d\mu(k) \lb\frac{1}{2}\lb A_{\sigma,k}^{2}-1\rb  - \ln{A_{\sigma,k}} \rb
\end{equation}
where expression for $I_{1}$, following transformations again in the discrete lattice case, is given by:
\begin{equation}
\begin{split}
    I_{1} &=  \int \mathcal{D}\sigma\lbq u\rbq \int d^{D}x g(x) U\lbq\bar\varphi(x) + \sum\limits_{\sigma}\int d\mu(k)\psi_{\sigma}(kx)A_{\sigma,k}u_{\sigma, k}\rbq \\
    &= \int d\varGamma e^{i\lambda \bar\varphi(x) - \frac{1}{2}\lambda^{2}\sum\limits_{\sigma}\int d\mu(k)A_{\sigma,k}^{2}\psi_{\sigma}^{2}(kx)}
\label{Cat_future}
\end{split}
\end{equation}
and, the function $A_{\sigma,k}$ is given in the standard way, again in the case of discrete lattice of functions,
\begin{equation}
    A_{\sigma,k}=\frac{1}{\sqrt{1+q_{\sigma}(k)}}.
\end{equation}
As a result, the desired expression for the majorant $\mathcal{F}$ on continuous lattice of functions is given by:
\begin{equation}
\boxed{
    \mathcal{F}\lbq g,\bar\varphi\rbq = \frac{1}{2}\sum\limits_{\sigma}\int d\mu(k)\lb\ln{\lb 1+q_{\sigma}(k)\rb} - \frac{q_{\sigma}(k)}
    {1+q_{\sigma}(k)} \rb + \int d\varGamma e^{i\lambda \bar\varphi(x) - \frac{1}{2}\lambda^{2}\sum\limits_{\sigma}\int d\mu(k)\frac{\psi_{\sigma}^{2}(kx)}{1+q_{\sigma}(k)}}}\, .
    \label{majoranta_Cat5}
\end{equation}
Having obtained the majorant $\mathcal{F}$ for the generating functional $\mathcal{G}$ on the continuous lattice of functions, it is important to note that the obtained majorant depends on pair of functions $q_{\sigma,k}$, with respect to which the majorant will be minimized in future. And, further derivations are possible only after specifying the theory; as a result, the polynomial theories $\varphi^{4}$ and $\varphi^{6}$, and the nonpolynomial sine-Gordon theory in $D$-dimensional spacetime, will both be considered in the following sections.
\section{Polynomial Theory $\varphi_{D}^{4}$}

\subsection{$\mathcal{S}$-Matrix of the $\varphi^{4}_{D}$ Theory in the Representation of Basis Functions (Discrete Lattice of Functions)}

In this section, we turn to the polynomial $\varphi^{4}$ and $\varphi^{6}$ theories in $D$-dimensional spacetime. We first consider the $\varphi^{4}$ theory in the case of discrete lattice of basis functions. We derive an explicit expression for the majorant of the generating functional $\mathcal{G}$, and then, in the approximation of the separable kernel, we obtain and solve the equations of the variational principle for the function $q_{s}$. Using the obtained solution, we derive the optimal majorant for $\mathcal{G}$. In similar way, we derive the expression for the $\varphi^{4}$ theory in the case of continuous lattice of basis functions. In the later part of the section, we give comment on vertex operators and the $\varphi^{6}$ theory.

For the $\varphi^{4}$ theory, the interaction Lagrangian $U$ is the polynomial function $U(\varphi)=\varphi^{4}$; field is set to zero i.e. $\bar\varphi=0$. In particular, the $\mathcal{S}$-matrix of the theory is given by
\begin{equation}
    \mathcal{S}[g] = 
    \lbc\prod\limits_{s}\int\limits_{-\infty}^{+\infty}\frac{dt_{s}}
{\sqrt{2\pi}}e^{-\frac{1}{2}t_{s}^{2}}\rbc e^{-\int d^{D}xg(x)
\left[ \sum\limits_{s}t_{s}D_{s}(x)\right]^{4}}.
\end{equation}
Writing, in fourth power of the field, the interaction action is represented by
\begin{equation}
    S_{1} = \sum\limits_{s^{(1)}}\sum\limits_{s^{(2)}}
    \sum\limits_{s^{(3)}}\sum\limits_{s^{(4)}}
    t_{s^{(1)}}t_{s^{(2)}}t_{s^{(3)}}t_{s^{(4)}} C_{s^{(1)}s^{(2)}s^{(3)}s^{(4)}}
\end{equation}
where coefficients $C_{s^{(1)}s^{(2)}s^{(3)}s^{(4)}}$ are given by 
\begin{equation}
    C_{s^{(1)}s^{(2)}s^{(3)}s^{(4)}} = \int d^{D}x g(x)D_{s^{(1)}}(x)D_{s^{(2)}}(x)D_{s^{(3)}}(x)D_{s^{(4)}}(x).
    \label{Cs1s2s3s4}
\end{equation}
In terms of the introduced notations, the $\mathcal{S}$-matrix of the theory is then given by
\begin{equation}
    \mathcal{S}[g] = 
    \lbc\prod\limits_{s}\int\limits_{-\infty}^{+\infty}\frac{dt_{s}}
{\sqrt{2\pi}}e^{-\frac{1}{2}t_{s}^{2}}\rbc e^{-\sum\limits_{s^{(1)}}\sum\limits_{s^{(2)}}
    \sum\limits_{s^{(3)}}\sum\limits_{s^{(4)}}
    t_{s^{(1)}}t_{s^{(2)}}t_{s^{(3)}}t_{s^{(4)}} C_{s^{(1)}s^{(2)}s^{(3)}s^{(4)}}}.
\end{equation}
The integral in the bracket, can be investigated by numerical methods, for instance, Monte Carlo methods; such research could be a topic for a separate paper. In this paper, however, we solve a simpler problem: we derive a majorant for $-\ln{\mathcal{S}}$, in the following sections.

\subsection{Majorant of the $\mathcal{S}$-Matrix of the $\varphi^{4}_{D}$ Theory (Discrete Lattice of Functions)}

Consider the expression (\ref{majoranta}) in relation to the theory with interaction $U(\varphi) = \varphi^{4}$ and zero field. The Fourier transform of interaction Lagrangian, here, is a distribution, given explicitly by 
\begin{equation}
    \Tilde{U}(\lambda) = 2\pi \frac{d^{4}\delta(\lambda)}{d\lambda^{4}}.
   \label{der4delta} 
\end{equation}
For the distribution (\ref{der4delta}), the integral over $x$ collapses into a coefficient $D_{s's''}$; and, the majorant (\ref{majoranta}), denoted $F$, is represented in the compact form
\begin{equation}
    F(q) \equiv 3 \sum\limits_{s',s''}\frac{D_{s's''}}{(1+q_{s'})(1+q_{s''})} + \frac{1}{2}\sum\limits_{s'}\lbq \ln{(1+q_{s'})-\frac{q_{s'}}{1+q_{s'}}}\rbq,
\end{equation}
where coefficients $D_{s's''}$ are given by
\begin{equation}
    D_{s's''}\equiv\int d^{D}x g(x) D^{2}_{s'}(x)D^{2}_{s''}(x).
\end{equation}
By the variational principle, we nee to find the function $q_{s}$ of the discrete variable $s$ (the set of numbers $q_{s}$) for which $F$ is extremal. In other words, we need to find the majorant closest to the original functional $\mathcal{G}$. To accomplish this, write the conditions for the extremum of $F$:
\begin{equation}
    \frac{\partial F(q)}{\partial q_{s}} = -6 \sum\limits_{s'}\frac{D_{ss'}}{(1+q_{s'})(1+q_{s})} + \frac{1}{2}\frac{q_{s}}{(1+q_{s})^{2}} = 0.
    \label{F_extr_gen} 
\end{equation}
By the condition (\ref{F_extr_gen}), the desired set of numbers $q_{s}$ satisfy the equation 
\begin{equation}
    q_{s} = 12 \sum\limits_{s'}\frac{D_{ss'}}{1+q_{s'}}.
    \label{q_min_nonsep}
\end{equation}
The resulting infinite system of nonlinear equations is a rather complex mathematical construction. We therefore need to simplify with the help of another inequality -- H\"{o}lder inequality, which allows an analogue of the system (\ref{q_min_nonsep}) to be obtained, but with separable kernel. Moreover, the H\"{o}lder inequality only increases the value of the majorant, making it quite legitimate. Besides, it makes no sense to solve a complex infinite system of nonlinear equations for the sake of the answer, which itself is a majorant.

\subsection{Integral Equation with Separable Kernel (Discrete Lattice of Functions)}

Consider the kernel $D_{s's''}$ in (\ref{q_min_nonsep}) and use the H\"{o}lder inequality. The following chain of equalities then holds:
\begin{equation}
\begin{split}
 D_{ss'} &= \int d^{D}xg(x)D_{s}^{2}(x)D_{s'}^{2}(x)= g\int d^{D}x\frac{g(x)}{g}D_{s}^{2}(x)D_{s'}^{2}(x) \leq \\ &g \sqrt{\int d^{D}x\frac{g(x)}{g}D_{s}^{4}(x)}
 \sqrt{\int d^{D}x\frac{g(x)}{g}D_{s'}^{4}(x)} \\&= \sqrt{\int d^{D}x g(x)D_{s}^{4}(x)}
 \sqrt{\int d^{D}xg(x)D_{s'}^{4}(x)}.
 \label{sep_kern_latt}
 \end{split}
\end{equation}
Let a measure be given by coupling constant $g(x)$ divided by its integral over the entire $D$-dimensional spacetime. The expression (\ref{sep_kern_latt}) is then given in the compact form:
\begin{equation}
    D_{ss'} \leq h_{s}h_{s'},\quad h_{s} = \sqrt{\int d^{D}xg(x)D_{s}^{4}(x)},\quad
    g=\int d^{D}xg(x).
\label{sep_kern_latt_comp}
\end{equation}
Substituting the separable kernel (\ref{sep_kern_latt_comp}) into equation (\ref{q_min_nonsep}), the equation for $q_{s}$, we then obtain the solution for $q_{s}$ in explicit analytical form:
\begin{equation}
    q_{s} = 12Ah_{s},\quad A = \sum\limits_{s}\frac{h_{s}}{1+12Ah_{s}}.
    \label{sep_kern_latt_sol}
\end{equation}
The solution actually is an imitation of the behavior of interaction constant in the lattice representation $h_{s}$ itself; the value $A$ is found from the second equality in (\ref{sep_kern_latt_sol}).

In the solution (\ref{sep_kern_latt_sol}), the majorant $F$ does not exceed its separable analog $F_{\mathrm{sep}}$; explicit expression of the inequality is given by
\begin{equation}
    F(q) \leq F_{\mathrm{sep}}(q)=
    \frac{1}{2}\sum\limits_{s}\ln{(1+12A h_{s})} - 3A^{2}.
\end{equation}
Plots of the obtained solution $q_{s}$ and the majorant $F_{\mathrm{sep}}$ is given constructed in the section devoted to numerical results. Note however, that all results obtained are valid for arbitrary dimension $D$ of $D$-dimensional spacetime corresponding to the lattice of basis functions.

\subsection{Majorant of the $\mathcal{S}$-Matrix of the $\varphi^{4}_{D}$ Theory (Continuous Lattice of Functions)}

In this subsection and the next, we implement the formulated procedure for the $\varphi^{4}$ theory in the case of continuous lattice of basis functions. Considering again the case of zero field $\bar\varphi= 0$, the expression (\ref{Cat_future}) for $I_{1}$ takes the form:
\begin{equation}
\begin{split}
    I_{1} &= \int\limits_{-\infty}^{+\infty}d\lambda\, \frac{d^{4}\delta(\lambda)}{d\lambda^{4}}\,e^{-\frac{1}{2}\lambda^{2}\sum\limits_{\sigma}\int d\mu(k) \frac{\psi_{\sigma}^{2}(kx)}{1+q_{\sigma}(k)}} \\
    &=3\sum\limits_{\sigma,\sigma'} \int d\mu(k)\int d\mu(k') \int d^{D}x g(x) \frac{\psi_{\sigma}^{2}\lb kx\rb\psi_{\sigma'}^{2}(k'x)}{\lb 1+q_{\sigma}(k)\rb\lb 1+q_{\sigma'}(k')\rb}
    \label{I_1}.
\end{split}
\end{equation}
As in the case of discrete lattice of basis functions, apply the H\"{o}lder inequality for split dependence on pairs of variables $(k,\sigma)$ and $(k',\sigma')$:
\begin{equation}
     \int d^{D}x g(x)\psi_{\sigma}^{2}\lb kx\rb\psi_{\sigma'}^{2}(k'x) \leq \sqrt{ \int d^{D}x g(x)\psi_{\sigma}^{4}\lb kx\rb }\sqrt{ \int d^{D}x g(x)\psi_{\sigma'}^{4}\lb k'x\rb }.
\end{equation}
Introduce again the notation $h_{\sigma}(k)$ for interaction constant in the lattice representation as follows:
\begin{equation}
h_{\sigma}(k) \equiv \sqrt{\int d^{D}x g(x) \psi_{\sigma}^{4}(kx)}.
\end{equation}
That is, the expression (\ref{I_1}) is majorized by a separable analog
\begin{equation}
    I_{1} \leq 3 \lb \sum\limits_{\sigma}\int d\mu(k) \frac{h_{\sigma}(k)}{1+q_{\sigma}(k)} \rb ^{2}.
\end{equation}
As usual, the desired majorant $\mathcal{F}$ does not exceed its separable analog $\mathcal{F}_{\mathrm{sep}}$; explicit expression for the inequality is given by:
\begin{equation}
\begin{split}
    &\mathcal{F}\lbq q \rbq \leq 
    \mathcal{F}_{\mathrm{sep}}\lbq q \rbq = \\
    &= \frac{1}{2}\sum\limits_{\sigma}\int d\mu(k) \lb \ln{\lb1+q_{\sigma}(k)\rb} - \frac{q_{\sigma}(k)}{1+q_{\sigma}(k)} \rb + 3 \lb \sum\limits_{\sigma}\int d\mu(k) \frac{h_{\sigma}(k)}{1+q_{\sigma}(k)}\rb^{2}
    \label{F_sep}.
\end{split}
\end{equation}
For the majorant $\mathcal{F}_{\mathrm{sep}}$ in (\ref{F_sep}), the extremum equation will be obtained in the next subsection. And, explicit expression for $\mathcal{F}_{\mathrm{sep}}$ on solving the equation for pair of functions $q_{\sigma,k}$ will be derived.

\subsection{Integral Equation with Separable Kernel (Continuous Lattice of Functions)}

From the extremum condition for the majorant $\mathcal{F}$ in (\ref{F_sep})
\begin{equation}
    \frac{\delta \mathcal{F}\lbq q \rbq}{\delta q_{\sigma}(k)} = 0
\end{equation}
equations for pair of functions $q_{\sigma}(k)$ are obtainable. At the same time, solutions of extremum equations for $\mathcal{F}_{\mathrm{sep}}$, being solutions of integral equations with separable kernels, are given by an analytical form
\begin{equation}
    q_{\sigma}(k) = 12Ah_{\sigma}(k).
    \label{q_k}
\end{equation}
As in the case of discrete lattice of functions, the solution imitates the behavior of interaction constant in the lattice representation $h_{\sigma}(k)$ itself; an explicit expression for the constant $A$ is given as follows:
\begin{equation}
    A = \sum\limits_{\sigma}\int d\mu(k) \frac{h_{\sigma}(k)}{1+q_{\sigma}(k)}.
    \label{A}
\end{equation}
Substituting the function (\ref{q_k}) into definition (\ref{A}), the constant $A$ is explicitly given by
\begin{equation}
    A = \sum\limits_{\sigma}\int d\mu(k) \frac{h_{\sigma}(k)}{1+12Ah_{\sigma}(k)}.
    \label{eq_for_A}
\end{equation}
By the solution (\ref{q_k}) -- (\ref{eq_for_A}), the desired majorant $\mathcal{F}_{\mathrm{sep}}$ now has the form
\begin{equation}
    \mathcal{F}_{\mathrm{sep}} \lbq q \rbq =   \frac{1}{2}\sum\limits_{\sigma}\int d\mu(k)\ln{\lb 1+12 Ah_{\sigma}(k)\rb}-3A^{2}.
\end{equation}
As usual, all obtained results are valid for arbitrary dimension $D$ of $D$-dimensional spacetime corresponding to the lattice of basis functions

There are a number of points to recollect in this section. First of all, the case of a nonzero field $\bar\varphi$ is investigated using methods developed in this section. Such field is represented as the series expansion over the functions $D_{s}$ with some coefficients $j_{s}$ on the discrete lattice. An arbitrary set of coefficients $j_{s}$ corresponds to an arbitrary field $\bar\varphi$. If the coefficients are chosen as sums of functions $D_{s}$ at fixed spacetime points $x_{\alpha}$ ($\alpha=1,2,...$) with coefficients $j_{\alpha}$, such choice of field corresponds to so-called vertex operators. As an illustration, the following equalities are given:
\begin{equation}
    \bar\varphi(x)=\sum\limits_{s}j_{s}D_{s}(x),\quad j_{s}=\sum\limits_{\alpha}j_{\alpha}D_{s}(x_{\alpha}).
    \label{vert_operator}
\end{equation}
Vertex operators play important role as objects, since the partial derivatives with respect to $j_{\alpha}$ of these operators give the values of the Green functions at the points $x_{\alpha}$. That is, instead of developing technique of inequalities for Green functions, one can consider the technique developed for the $\mathcal{S}$-matrix of the theory with a field (\ref{vert_operator}). The research may be the subject of a separate paper.

Finally, a few words on the polynomial $\varphi^{6}$ theory (for arbitrary $D$): For the theory, the Fourier transform of the interaction Lagrangian is also the distribution, explicit form of which is given by:
\begin{equation}
    \Tilde{U}(\lambda) = -2\pi \frac{d^{6}\delta(\lambda)}{d\lambda^{6}}.
   \label{der6delta} 
\end{equation}
All the methods developed in this section allow investigation of such field theory by actually following the procedure already done. In particular, the majorants $F_{\mathrm{sep}}$ and $\mathcal{F}_{\mathrm{sep}}$ will be constructed on discrete and continuous lattices of basis functions, respectively. Such research may also be the subject of a separate paper.
\section{Nonpolynomial Sine-Gordon Theory}

\subsection{Majorant of the $\mathcal{S}$-Matrix of the Sine-Gordon Theory (Discrete Lattice of Functions)}

In this section, we consider the nonpolynomial sine-Gordon theory in the $D$-dimensional spacetime. First we consider the theory in the case of the discrete lattice of basis functions. We derive an explicit expression for the majorant of the generating functional $\mathcal{G}$, and then we obtain the equations of the variational principle for the function $q_{s}$. Next, we find the upper bound $q^{(+)}_{s}$ and the lower bound $q^{(-)}_{s}$ for the function $q_{s}$. Then we find a simplified majorant for the generating functional $\mathcal{G}$, for which we obtain and solve the equations of the variational principle for the corresponding function $q_{s}$ (we denote it by the same symbol, since its position in the text is such that will not confuse). Using the obtained solution, we define the optimal majorant for $\mathcal{G}$ in the simplified case. Then we will repeat the formulated program for the sine-Gordon theory in the case of the continuous lattice of basis functions.

Consider the general expression for majorant (\ref{majoranta}) in application to the sine-Gordon theory. In this case, the Fourier transform of the interaction Lagrangian $\tilde{U}$ is the distribution:
\begin{equation}
    \tilde{U}(\lambda) = \pi\lbq\delta(\lambda-\eta)+\delta(\lambda+\eta)\rbq,
\label{for_scale_disc}
\end{equation}
and the majorant takes the following form
\begin{equation}
 F(q) \equiv \int d^{D}x g(x)\cos{\lbq\eta\bar\varphi(x)\rbq}e^{- \frac{1}{2}\eta^{2} \sum\limits_{s}\frac{D_{s}^{2}(x)}{{1+q_{s}}}}+\frac{1}{2}\sum\limits_{s}\left[\ln{(1+q_{s})}-\frac{q_{s}}{1+q_{s}}\right].
 \label{sinG_dislat1}
\end{equation}
The expression (\ref{sinG_dislat1}) depends on the field $\bar\varphi$. We set to zero the variational derivative of $F$ over the field $\bar\varphi$ in order to find the condition for the extremal field, therefore, the vacuum of the sine-Gordon theory. The sought expression reads as follows:
\begin{equation}
    \frac{\delta F(q)}{\delta \bar\varphi (y)} = -\eta\int d^{D}x g(x)\sin{\lbq\eta\bar\varphi(x)\rbq}e^{- \frac{1}{2}\eta^{2} \sum\limits_{s'}\frac{{D_{s'}^{2}(x)}}{{1+q_{s'}}}}\delta^{(D)}(x-y) = 0,
\end{equation}
which gives us the condition $\eta\bar\varphi(y) = \pi m, m\in \mathbb{Z}$. Let us choose the (vacuum) value of the field $\eta\bar\varphi_{0}=\pi(2m+1)$. In this case, we reach the global minimum of the sine-Gordon theory. In the future, the dependence on $\bar\varphi_{0}$ will not be explicitly indicated. The expression for majorant in this case is:
\begin{equation}
    F(q) = -\int d^{D}x g(x)e^{- \frac{1}{2}\eta^{2} \sum\limits_{s}\frac{D_{s}^{2}(x)}{{1+q_{s}}}}+\frac{1}{2}\sum\limits_{s}\left[\ln{(1+q_{s})}-\frac{q_{s}}{1+q_{s}}\right].
\label{sinG_dislat_for_simple1}
\end{equation}
From the minimum condition (the partial derivatives of $F$ with respect to $q_{s}$ equal to zero), we obtain the equations of the variational principle for the set $q_{s}$, for which $F$ is minimal:
\begin{equation}
    \frac{\partial F(q)}{\partial q_{s}} = 0,\quad q_{s} = \eta ^{2}\int d^{D}x g(x)e^{- \frac{1}{2}\eta^{2} \sum\limits_{s'}\frac{{D_{s'}^{2}(x)}}{{1+q_{s'}}}}D_{s}^{2}(x).
\label{sinG_dislat2}
\end{equation}
The resulting infinite system of nonlinear equations is a rather complex mathematical construction. The situation is even more complicated than in the $\varphi^{4}$ theory, since the system (\ref{sinG_dislat2}) contains an integral over $x$ in an explicit form. For this reason, we will find upper bound $q^{(+)}_{s}$ and lower bound $q^{(-)}_{s}$ for the function $q_{s}$, in other words, such functions, that $q^{(-)}_{s}\leq q_{s}\leq q^{(+)}_{s}$.

\subsection{Estimates for the Solution of the Variational Principle Equations (Discrete Lattice of Functions)}

\subsubsection{Upper Bound}

We begin by defining the upper bound $q^{(+)}_{s}$. We use the fact that the natural measure can be defined on the discrete lattice, since the following equality holds:
\begin{equation}
    \sum\limits_{s}D_{s}^{2}(x) = G(0).
\end{equation}
Next, in the expression (\ref{sinG_dislat2}) we use the Jensen inequality for a convex function which is the exponential function in this case. Then for the function $q_{s}$ in (\ref{sinG_dislat2}) the following is true:
\begin{equation}
\begin{split}
    &q_{s} \leq \frac{\eta^{2}}{G(0)}\int d^{D}x g(x)D_{s}^{2}(x)\sum\limits_{s'}D_{s'}^{2}(x)e^{- \frac{1}{2}\eta^{2}G(0) \frac{1}{{1+q_{s'}}}} \leq \\& \leq \frac{\eta ^{2}}{G(0)}\sum\limits_{s'}D_{ss'}e^{-\frac{1}{2} \eta^{2}G(0)\frac{1}{{1+q_{s'}}}},\quad 
    D_{ss'}\equiv\int d^{D}x g(x)D_{s}^{2}(x)D_{s'}^{2}(x).
\label{q+}
\end{split}
\end{equation}
Note that in the expression (\ref{q+}) the integral over $x$ appears only in the definition of the coefficient $D_{ss'}$.

We further assume that the coefficient $D_{ss'}$ appearing in the expression (\ref{q+}) is separable -- thus, we will only intensify the inequality (\ref{q+}). We use the H\"{o}lder inequality to split the coefficient $D_{ss'}$. The result repeats the case of the $\varphi^{4}$ theory:
\begin{equation}
    D_{ss'} \leq h_{s}h_{s'},\quad h_{s}\equiv\sqrt{\int d^{D}x g(x)D_{s}^{4}(x)}.
\label{sinG_dislat_sep_ker}
\end{equation}
Substituting the splitting (\ref{sinG_dislat_sep_ker}) into the expression (\ref{q+}), and closing it with the function $q^{(+)}_{s}$, we obtain the equation with the separable kernel for the upper bound $q^{(+)}_{s}$. The solution of this equation has a simple analytical form:
\begin{equation}
    q_{s} \leq q^{(+)}_{s} \equiv \frac{\eta^{2}A^{(+)}h_{s}}{G(0)}, 
    \quad A^{(+)} = \sum\limits_{s}h_{s}e^{-\frac{1}{2} \frac{\eta^{2}G^{2}(0)}{G(0)+\eta^{2}A^{(+)}h_{s}}}.
\label{sinG_dislat3}
\end{equation}
The first equality in the expression (\ref{sinG_dislat3}) shows that the estimate $q^{(+)}_{s}$ actually repeats the behavior of the interaction constant in the lattice representation $h_{s}$ itself. The second equality in the expression (\ref{sinG_dislat3}) gives the equation for determining the coefficient $A^{(+)}$. Such an equation can be solved by numerical methods.

\subsubsection{Lower Bound}

Let us proceed to the definition of the lower bound $q^{(-)}_{s}$. We first introduce the following definition for the $D_{s}$ harmonics of the interaction constant in the lattice representation:
\begin{equation}
    \int d^{D}x g(x)D_{s}^{2}(x) \equiv g^{(0)}_{s}.
\end{equation}
Now we can write the following chain of inequalities, based on the Jensen inequality for the exponential function and the H\"{o}lder inequality for splitting the arising coefficient $D_{ss'}$:
\begin{equation}
\begin{split}
    q_{s} = \frac{\eta ^{2}g^{(0)}_{s}}{g^{(0)}_{s}}\int d^{D}x g(x)D_{s}^{2}(x)e^{-\frac{1}{2}\eta^{2}
    \sum\limits_{s'}\frac{D_{s'}^{2}(x)}{{1+q_{s'}}}} &\geq \eta^{2}g^{(0)}_{s}e^{- \frac{\eta^{2}}{2g^{(0)}_{s}}\int d^{D}x g(x)D_{s}^{2}(x)\sum\limits_{s'}\frac{D_{s'}^{2}(x)}{{1+q_{s'}}}} \geq \\  &\geq \eta ^{2}g^{(0)}_{s}e^{- \frac{\eta^{2}}{2g^{(0)}_{s}}h_{s}\sum\limits_{s'}
    \frac{h_{s'}}{{1+q_{s'}}}}.
\end{split}
\label{sinG_dislat4}
\end{equation}
The expression (\ref{sinG_dislat4}) allows us to obtain the equation with the separable kernel for the lower bound $q^{(-)}_{s}$. The solution of this equation has a simple analytical form:
\begin{equation}
    q_{s}\geq q^{(-)}_{s} \equiv \eta ^{2}g^{(0)}_{s}e^{- \frac{\eta^{2}}{2g^{(0)}_{s}}A^{(-)}h_{s}}, \quad 
    A^{(-)} \equiv \sum\limits_{s}\frac{h_{s}}{1+ \eta^{2}g^{(0)}_{s}e^{-\frac{1}{2}\eta^{2}A^{(-)}
    \frac{h_{s}}{g^{(0)}_{s}}}}.
\label{sinG_dislat5}
\end{equation}
The second equality in the expression (\ref{sinG_dislat5}) gives an equation for determining the coefficient $A^{(-)}$. Such an equation can be solved by numerical methods.

Note that the estimates $q^{(+)}_{s}$ and $q^{(-)}_{s}$ are simpler for calculations than $q_{s}$, since explicit integration over $x$ in the corresponding expressions is missing. The plots of the obtained estimates $q^{(+)}_{s}$ and $q^{(-)}_{s}$ will be constructed in the next section.

\subsection{Simple Majorant of $\mathcal{G}$ on the Discrete Lattice of Functions}

In the final subsection devoted to the sine-Gordon theory on the discrete lattice of functions, we derive a simple majorant. First we introduce the measure of integration over to $x$:
\begin{equation}
    \frac{1}{g}\int d^{D}x g(x)(...),\quad g \equiv \int d^{D}x g(x).
\end{equation}
Next, we use Jensen inequality for the exponential function in the expression for the majorant (\ref{sinG_dislat_for_simple1}) itself. The following transformations are valid:
\begin{equation}
    -\int d^{D}x g(x)e^{-\frac{1}{2}\eta^{2}\sum\limits_{s}\frac{D_{s}^{2}(x)}{{1+q_{s}}}} \leq -ge^{-\frac{1}{2}\eta^{2}\frac{1}{g}\int d^{D}x g(x)\sum\limits_{s}\frac{D_{s}^{2}(x)}{{1+q_{s}}}} = -ge^{-\frac{1}{2}\eta^{2}\frac{1}{g}\sum\limits_{s}
    \frac{g^{(0)}_{s}}{{1+q_{s}}}}.
\label{sinG_dislat_for_simple2}
\end{equation}
The expression (\ref{sinG_dislat_for_simple2}) allows us to write the expression for the simple majorant $F_{\mathrm{simple}}$:
\begin{equation}
    F_{\mathrm{simple}}(q) = -ge^{-\frac{1}{2}\eta^{2}\frac{1}{g}\sum\limits_{s}\frac{g^{(0)}_{s}}{{1+q_{s}}}}+\frac{1}{2}\sum\limits_{s}\left[\ln{(1+q_{s})}
    -\frac{q_{s}}{1+q_{s}}\right].
\label{sinG_dislat_for_simple3}
\end{equation}
Next, we must derive and solve the equations of the variational principle for $F_{\mathrm{simple}}$ (\ref{sinG_dislat_for_simple3}).

The minimum condition is:
\begin{equation}
    \frac{\partial F_{\mathrm{simple}}(q)}{\partial q_{s}} =  \frac{1}{2}\frac{q_{s}}{(1+q_{s})^{2}} -\frac{\eta^{2}}{2} e^{-\frac{1}{2}\eta^{2}\frac{1}{g}\sum\limits_{s'}\frac{g^{(0)}_{s'}}{{1+q_{s'}}}}\frac{g^{(0)}_{s}}{(1+q_{s})^{2}} = 0.
\label{sinG_dislat_for_simple4}
\end{equation}
The expression (\ref{sinG_dislat_for_simple4}) allows us to obtain a simple equation for $q_{s}$. The analytical solution to this equation reads as follows:
\begin{equation}
    q_{s} = \eta^{2}g^{(0)}_{s}e^{-\frac{1}{2}\eta^{2}\frac{1}{g}
    A_{\mathrm{simple}}}, \quad 
    A_{\mathrm{simple}} = \sum\limits_{s} \frac{g^{(0)}_{s}}{1+\eta^{2}g^{(0)}_{s}
    e^{-\frac{1}{2}\eta^{2}\frac{1}{g}A_{\mathrm{simple}}}}.
\label{sinG_dislat_for_simple5}
\end{equation}
As always, the value $A_{\mathrm{simple}}$ is determined by the second equality in (\ref{sinG_dislat_for_simple5}). Substituting then the solution $q_{s}$ in the form (\ref{sinG_dislat_for_simple5}) into the expression (\ref{sinG_dislat_for_simple3}) for $F_{\mathrm{simple}}$, we obtain the desired simple majorant on the solution:
\begin{equation}
    F_{\mathrm{simple}}(q) = \frac{1}{2} \sum\limits_{s} \ln{(1+q_{s})} - \left(g+\frac{\eta^{2}A_{\mathrm{simple}}}{2}\right)
    e^{-\frac{1}{2}\eta^{2}\frac{1}{g}A_{\mathrm{simple}}}.
\label{sinG_dislat_for_numcalc}
\end{equation}
This concludes the presentation of the sine-Gordon theory on the discrete lattice of functions. In the remaining subsections of this section, we consider the sine-Gordon theory on the continuous lattice of functions.

\subsection{Majorant of the $\mathcal{S}$-Matrix of the Sine-Gordon Theory (Continuous Lattice of Functions)}

In this and the next subsections, we implement the formulated program for the sine-Gordon theory in the case of the continuous lattice of basis functions. Let us consider the expression (\ref{majoranta_Cat5}) in application to the sine-Gordon theory. In this case, the expression (\ref{majoranta_Cat5}) reads:
\begin{equation}
\begin{split}
    \mathcal{F}\lbq q\rbq &\equiv \frac{1}{2} \sum\limits_{\sigma}\int d\mu(k) \lb \ln{(1+q_{\sigma}(k))} - \frac{q_{\sigma}(k)}{1+q_{\sigma}(k)} \rb + \\ &+ \int d^{D}x g(x)\cos{\lbq\eta\bar\varphi(x)\rbq} e^{- \frac{1}{2}\eta^{2}\sum\limits_{\sigma}\int d\mu(k)\frac{\psi_{\sigma}^{2}(kx)}{1+q_{\sigma}(k)}}.
\end{split}
\label{sinG_contlat_Cat1}
\end{equation}
The minimum of the functional (\ref{sinG_contlat_Cat1}) is obtained when $\eta\bar\varphi(x)=\eta\bar\varphi_{0}=\pi(2m+1), m\in \mathbb{Z}$ (sine-Gordon theory vacuum). In the future, the dependence on $\bar\varphi_{0}$ will not be explicitly indicated. The expression for the majorant (\ref{sinG_contlat_Cat1}) in this case has the following form:
\begin{equation}
\begin{split}
    \mathcal{F}\lbq q\rbq = \frac{1}{2} \sum\limits_{\sigma}\int d\mu(k)\lb \ln{(1+q_{\sigma}(k))} -\frac{q_{\sigma}(k)}{1+q_{\sigma}(k)} \rb - \int d^{D}x g(x) e^{- \frac{1}{2}\eta^{2}\sum\limits_{\sigma}\int d\mu(k)\frac{\psi_{\sigma}^{2}(kx)}{1+q_{\sigma}(k)}}.
\end{split}
\label{sinG_contlat_Cat2}
\end{equation}
From the minimum condition of the functional (\ref{sinG_contlat_Cat2}), we obtain the following equation of the variational principle for the pair of functions $q_{\sigma}(k)$:
\begin{equation}
    \frac{\delta \mathcal{F}\lbq q\rbq}
    {\delta q_{\sigma}(k)} = 0,\quad q_{\sigma}(k) = \eta^{2}\int d^{D}x g(x)\psi_{\sigma}^{2}(kx) e^{- \frac{1}{2}\eta^{2}\sum\limits_{\sigma'}\int d\mu(k')\frac{\psi_{\sigma'}^{2}(k'x)}{1+q_{\sigma'}(k')}}.
\label{sinG_contlat_Cat3}
\end{equation}
The equation (\ref{sinG_contlat_Cat3}) is an analogue of the system of equations (\ref{sinG_dislat2}) used in the discrete case. As on the discrete lattice, to determine the solution $q_{\sigma}(k)$, it suffices to find upper bound $q^{(+)}_{\sigma}(k)$ and lower bound $q^{(-)}_{\sigma}(k)$.

\subsection{Estimates for the Solution of the Variational Principle Equations (Continuous Lattice of Functions)}

\subsubsection{Upper Bound}

We begin with the derivation of the upper bound $q^{(+)}_{\sigma}(k)$. We use the fact that the natural measure can be defined on the continuous lattice, since the following equality holds (a consequence from the explicit expression for the propagator of the theory):
\begin{equation}
    \sum\limits_{\sigma}\int d\mu(k)\psi_{\sigma}^{2}(kx) = 1.
\end{equation}
Then we use again the Jensen inequality for the exponential function:
\begin{equation}
    e^{- \frac{1}{2}\eta^{2}\sum\limits_{\sigma}\int d\mu(k)\frac{\psi_{\sigma}^{2}(kx)}{1+q_{\sigma}(k)}} \leq \sum\limits_{\sigma}\int d\mu(k)\psi_{\sigma}^{2}(kx)e^{- \frac{1}{2}\eta^{2}\frac{1}{1+q_{\sigma}(k)}}.
    \label{sin_major}
\end{equation}
Given the inequality (\ref{sin_major}), we obtain the following expression for the pair of functions $q_{\sigma}(k)$:
\begin{equation}
    q_{\sigma}(k) \leq \eta^{2}\sum\limits_{\sigma'}\int d\mu(k')e^{- \frac{1}{2}\eta^{2}\frac{1}{1+q_{\sigma'}(k')}} \int d^{D}x g(x)\psi_{\sigma}^{2}(kx)\psi_{\sigma'}^{2}(k'x).
    \label{q0sigma}
\end{equation}
As in the case of the $\varphi^{4}$ theory, we use the H\"{o}lder inequality to split the dependence on pairs of the variables $(k,\sigma)$ and $(k',\sigma')$:
\begin{equation}
    \int d^{D}x g(x)\psi_{\sigma}^{2}(kx)\psi_{\sigma'}^{2}(k'x) \leq h_{\sigma}(k)h_{\sigma'}(k'),\quad 
    h_{\sigma}(k) \equiv \sqrt{\int d^{D}x g(x)\psi_{\sigma}^{4}(kx)}.
\label{sinG_contlat_Cat4}
\end{equation}
The expression (\ref{sinG_contlat_Cat4}) introduces the notation $h_{\sigma}(k)$ for the interaction constant in the lattice representation.

Substituting the splitting (\ref{sinG_contlat_Cat4}) into the expression (\ref{q0sigma}), and closing it with the function $q^{(+)}_{\sigma}(k)$, we obtain the equation with the separable kernel for the upper bound $q^{(+)}_{\sigma}(k)$. The solution of this equation has a simple analytical form:
\begin{equation}
     q_{\sigma}(k) \leq q^{(+)}_{\sigma}(k) = 
     \eta^{2}A^{(+)}h_{\sigma}(k),\quad 
     A^{(+)} = \sum\limits_{\sigma}\int d\mu(k) h_{\sigma}(k) 
     e^{-\frac{1}{2}\frac{\eta^{2}}{1+\eta^{2}A^{(+)}h_{\sigma}(k)}}.
\label{sinG_contlat_Cat5}
\end{equation}
The first equality in the expression (\ref{sinG_contlat_Cat5}) shows that the estimate $q^{(+)}_{\sigma}(k)$ actually repeats the behavior of the interaction constant in the lattice representation $h_{\sigma}(k)$ itself. The second equality in the expression (\ref{sinG_contlat_Cat5}) gives the equation for determining the coefficient $A^{(+)}$.

\subsubsection{Lower Bound}

Let us proceed to the derivation of the lower bound $q^{(-)}_{\sigma}(k)$. First we introduce the following definition for the $(k,\sigma)$ harmonics of the interaction constant in the lattice representation:
\begin{equation}
    \int d^{D}x g(x)\psi_{\sigma}^{2}(kx) \equiv g^{(0)}_{\sigma}(k).
\end{equation}
Now we can write the following chain of inequalities based on Jensen and H\"{o}lder inequalities:
\begin{equation}
\begin{split}
    q_{\sigma}(k) &= \frac{\eta^{2}g^{(0)}_{\sigma}(k)}{g^{(0)}_{\sigma}(k)}
    \int d^{D}x g(x)\psi_{\sigma}^{2}(kx) e^{-\frac{1}{2}\eta^{2}\sum\limits_{\sigma'}\int d\mu(k')\frac{\psi_{\sigma'}^{2}(k'x)}{1+q_{\sigma'}(k')}} \geq \\ &\geq \eta^{2}g^{(0)}_{\sigma}(k) 
    e^{-\frac{1}{2}\eta^{2}\frac{1}{g^{(0)}_{\sigma}(k)} \sum\limits_{\sigma'}\int d\mu(k')\frac{1}{1+q_{\sigma'}(k')}\int d^{D}x g(x)\psi_{\sigma}^{2}(kx)\psi_{\sigma'}^{2}(k'x)} \geq \\  &\geq \eta^{2}g^{(0)}_{\sigma}(k) 
    e^{-\frac{1}{2}\eta^{2}\frac{h_{\sigma}(k)}{g^{(0)}_{\sigma}(k)} \sum\limits_{\sigma'}\int d\mu(k')\frac{h_{\sigma'}(k')}
    {1+q_{\sigma'}(k')}}.
\end{split}
\label{sinG_contlat_Cat6}
\end{equation}
The expression (\ref{sinG_contlat_Cat6}) allows us to obtain the equation with the separable kernel for the lower bound $q^{(-)}_{\sigma}(k)$. The solution of this equation has a simple analytical form:
\begin{equation}
    q_{\sigma}(k) \geq q^{(-)}_{\sigma}(k) = 
    \eta^{2}g^{(0)}_{\sigma}(k) e^{-\frac{1}{2}\eta^{2}A^{(-)}
    \frac{h_{\sigma}(k)}{g^{(0)}_{\sigma}(k)}}.
\label{sinG_contlat_Cat7}
\end{equation}
The value $A^{(-)}$ appearing in the expression (\ref{sinG_contlat_Cat7}) is determined by the following equation:
\begin{equation}
\begin{split}
    A^{(-)} &= \sum\limits_{\sigma}\int d\mu(k)\frac{h_{\sigma}(k)}
    {1+\eta^{2}g^{(0)}_{\sigma}(k)e^{-\frac{1}{2}\eta^{2}A^{(-)}
    \frac{h_{\sigma}(k)}{g^{(0)}_{\sigma}(k)}}}.
\end{split}
\label{sinG_contlat_Cat8}
\end{equation}
At the end of the subsection, we note that, as in the case of the discrete lattice, the estimates $q^{(+)}_{\sigma}(k)$ and $q^{(-)}_{\sigma}(k)$ are easier to compute than $q_{\sigma}(k)$, since there is no explicit integration over $x$ in the corresponding expressions. It remains to derive the simple majorant of $\mathcal{G}$ on the continuous lattice of functions to which the next final subsection is devoted.

\subsection{Simple Majorant of $\mathcal{G}$ on the Continuous Lattice of Functions}

Similarly to the case of simple majorant on the discrete lattice at the beginning, we use the Jensen inequality for the exponent function in the expression for the majorant (\ref{sinG_contlat_Cat2}) itself. The following transformations are valid:
\begin{equation}
\begin{split}
    -\int d^{D}x g(x)e^{-\frac{1}{2}\eta^{2}\sum\limits_{\sigma}\int d\mu(k)\frac{\psi_{\sigma}^{2}(kx)}{1+q_{\sigma}(k)}} &\leq -g e^{-\frac{1}{2}\eta^{2}\frac{1}{g}\sum\limits_{\sigma}\int d\mu(k)\frac{1}{1+q_{\sigma}(k)}\int d^{D}x g(x)\psi_{\sigma}^{2}(kx)} = \\ &= -g e^{-\frac{1}{2}\eta^{2}\frac{1}{g}\sum\limits_{\sigma}\int d\mu(k)\frac{g_{\sigma}^{(0)}(k)}{1+q_{\sigma}(k)}}.
\end{split}
\label{sinG_contlat_Cat9}
\end{equation}
The expression (\ref{sinG_contlat_Cat9}) allows us to write the expression for the simple majorant $\mathcal{F}_{\mathrm{simple}}$:
\begin{equation}
    \mathcal{F}_{\mathrm{simple}}\lbq q\rbq = \frac{1}{2} \sum\limits_{\sigma}\int d\mu(k) \lb \ln{(1+q_{\sigma}(k))} - \frac{q_{\sigma}(k)}{1+q_{\sigma}(k)} \rb - 
    g e^{-\frac{1}{2}\eta^{2}\frac{1}{g}\sum\limits_{\sigma}\int d\mu(k)\frac{g_{\sigma}^{(0)}(k)}{1+q_{\sigma}(k)}}.
\label{sinG_contlat_Cat10}
\end{equation}
Next, we must derive and solve the equations of the variational principle for $\mathcal{F}_{\mathrm{simple}}$ (\ref{sinG_contlat_Cat10}).

From the minimum condition, we find the pair of functions $q_{\sigma}(k)$, for which the simple majorant is minimal, in analytic form:
\begin{equation}
    \frac{\delta \mathcal{F}_{\mathrm{simple}}\lbq q\rbq}
    {\delta q_{\sigma}(k)} = 0, \quad q_{\sigma}(k) = \eta^{2}g_{\sigma}^{(0)}(k)e^{-\frac{1}{2}\eta^{2}
    \frac{1}{g}A_{\mathrm{simple}}}.
\label{sinG_contlat_Cat11}
\end{equation}
The value $A^{(+)}$ appearing in the expression (\ref{sinG_contlat_Cat11}) is determined by the following equation:
\begin{equation}
\begin{split}
    A_{\mathrm{simple}} &= \sum\limits_{\sigma}\int d\mu(k) \frac{g_{\sigma}^{(0)}(k)}{1 + \eta^{2}g_{\sigma}^{(0)}(k)e^{-\frac{1}{2}\eta^{2}\frac{1}{g}
    A_{\mathrm{simple}}}}.
\end{split}
\label{sinG_contlat_Cat12}
\end{equation}
Substituting then the solution $q_{\sigma}(k)$ in the form (\ref{sinG_contlat_Cat11}) -- (\ref{sinG_contlat_Cat12}) into the expression (\ref{sinG_contlat_Cat10}) for $\mathcal{F}_{\mathrm{simple}}$, we obtain the desired simple majorant on the solution:
\begin{equation}
\begin{split}
    \mathcal{F}_{\mathrm{simple}}\lbq q\rbq &= \frac{1}{2} \sum\limits_{\sigma}\int d\mu(k) \ln{(1+q_{\sigma}(k))} -\left(g+\frac{\eta^{2}A_{\mathrm{simple}}}{2}\right)
    e^{-\frac{1}{2}\eta^{2}\frac{1}{g}A_{\mathrm{simple}}}.
\end{split}
\end{equation}
This concludes the presentation of the sine-Gordon theory on the continuous lattice of functions. In conclusion, we note that the $\mathcal{S}$-matrix of this theory can be investigated by the methods of statistical physics, if we write $\mathcal{S}$ in the form of the grand canonical partition function. In this paper, the sine-Gordon theory was chosen as an example to demonstrate the methods developed in the framework of the general approach based on the representation of the $\mathcal{S}$-matrix of an arbitrary QFT in terms of basis functions. The study of the sine-Gordon theory in $D$-dimensional spacetime by the methods of statistical physics can be the subject of a separate paper.
\section{From $D=1$ to $D=26$ and Beyond}

\subsection{Polynomial Theory $\varphi_{1}^{4}$}

In this subsection, we consider the majorant $F_{\mathrm{sep}}$ for the $\mathcal{S}$-matrix of the polynomial $\varphi^{4}_{D}$ theory on the discrete lattice of basis functions for a particular type of the coupling constant $g(x)$ and the propagator $D(x)$ in the dimension $D=1$. Both functions will be chosen in the form of the Gaussian function. Such a choice allows us to simplify analytical calculations in the maximal way, since the integrals of the product of the Gaussian function and the Hermite functions are again expressed through themselves, as it is well known from the theory of the Gauss--Weierstrass integral transformation \cite{bateman1981functions}. This happens, for example, for functions $D_{s}(x)$.

\subsubsection{Analytics}

Let us choose the coupling constant $g(x)$ as a Gaussian function with amplitude $g(0)$ and (inverse) width $w$:
\begin{equation}
    g(x) = g(0)e^{-\frac{1}{2}wx^2}.
\end{equation}
The parameter $w$ determines how fast the interaction falls, therefore, must be expressed through some infrared length $L$. In the realistic model, the value of $L$ should be the largest of all possible lengths in the system.

Next, we consider the propagator of the theory $D(x)$. We also select it as a Gaussian function with amplitude $D(0)$ and width $\varepsilon$. Note that the case of the local QFT for such a propagator is obtained by passing to the limit $\varepsilon\rightarrow 0$ when choosing the propagator amplitude $D(0)\sim 1/\sqrt{\varepsilon}$. Within the framework of the nonlocal QFT, the expression for the propagator with finite $\varepsilon$ reads:
\begin{equation}
    D(x) =D(0) e^{-\frac{1}{2\varepsilon}x^2}.
\end{equation}
In the process of calculating the majorant, we need to calculate the functions $D_{s}(x)$:
\begin{equation}
D_{s}(x) =\frac{D(0)}{\sqrt{2^{s}s!\sqrt{\pi}}} \int dz e^{-(x-z)^{2}}  H_{s}(z)e^{-\frac{1}{2}z^2}.
\label{num:Ds}
\end{equation}
To calculate the integral in expression (\ref{num:Ds}), we use the following result from the theory of the Gauss--Weierstrass integral transformation \cite{bateman1981functions} (the function $F_{u}$ is an image of the integral transformation):
\begin{equation}
\begin{split}
    F_{u}(z) \equiv  \int\limits_{-\infty}^{+\infty}dy H_{n}(y)e^{-\frac{1}{2u}(z-y)^{2}} = \sqrt{2\pi u}\sqrt{(1-2u)^{n}} H_{n}\lb\frac{z}{\sqrt{1-2u}}\rb,\quad 0 \leq u < \frac{1}{2}.
\label{weierstrass1}
\end{split}
\end{equation}
For our problem, the application of the expression (\ref{weierstrass1}) reads:
\begin{equation}
\begin{split}
    \int\limits_{-\infty}^{+\infty}dy H_{n}(y)e^{-\frac{1}{2}y^{2}}e^{-\frac{1}{2\varepsilon}(x-y)^{2}} = e^{-\frac{1}{2}\frac{x^{2}}{1+\varepsilon}}F_{\frac{\varepsilon}{1+\varepsilon}}\lb \frac{x}{1+\varepsilon}\rb.
\label{weierstrass2}
\end{split}
\end{equation}
Using the expression (\ref{weierstrass2}), we can calculate the integral for the functions $D_{s}(x)$ (\ref{num:Ds}). After substituting the image $F_{u}$, the result reads as follows:
\begin{equation}
    D_{s}(x) =\frac{D(0)}{\sqrt{2^{s}s!\sqrt{\pi}}}\sqrt{\frac{2\pi\varepsilon}{1+\varepsilon}\lb\frac{1-\epsilon}{1+\epsilon}\rb^{s}} H_{s}\lb x\sqrt{\frac{1+\varepsilon}{1-\varepsilon}}\rb e^{-\frac{1}{2}\frac{x^{2}}{1+\varepsilon}}.
\end{equation}
In this form, the functions $D_{s}(x)$ will be used in the numerical calculations.

The integrals for the coupling constant in the lattice representation $h_{s} = \sqrt{\int dx g(x)D_{s}^{4}(x)}$ are calculated numerically. Having determined $h_{s}$, we use them to find the value of the constant $A$ from the following equation:
\begin{equation}
    A = \sum\limits_{s=0}^{\infty}\frac{h_{s}}{1+12A h_{s}}.
\end{equation}
Thus, having determined the quantities $h_{s}$ and $A$, we arrive at the final expression for the majorant $F_{\mathrm{sep}}$, which is easily found numerically:
\begin{equation}
    F_{\mathrm{sep}} = \frac{1}{2}\sum\limits_{s=0}^{\infty}\ln{\lb1+12A h_{s}\rb}-3A^{2}.
    \label{num:maj}
\end{equation}
In the following subsection the results of the calculation on the lattice $s_{\mathrm{max}}=20$ at various values of the model parameters are presented. Let us note that the results obtained on the lattice $s_{\mathrm{max}}=10$ differ from those given in the present paper insignificantly.

\subsubsection{Plots}

In this subsection, we present numerical results (plots) for the majorant $F_{\mathrm{sep}}$ for the $\mathcal{S}$-matrix of the polynomial theory $\varphi^{4}_{1}$ on the discrete lattice of basis functions. We also present plots of the variation function $q_{s}$.

\begin{figure}[H]
\centering
\begin{minipage}{.5\textwidth}
  \centering
  \includegraphics[scale=0.26]{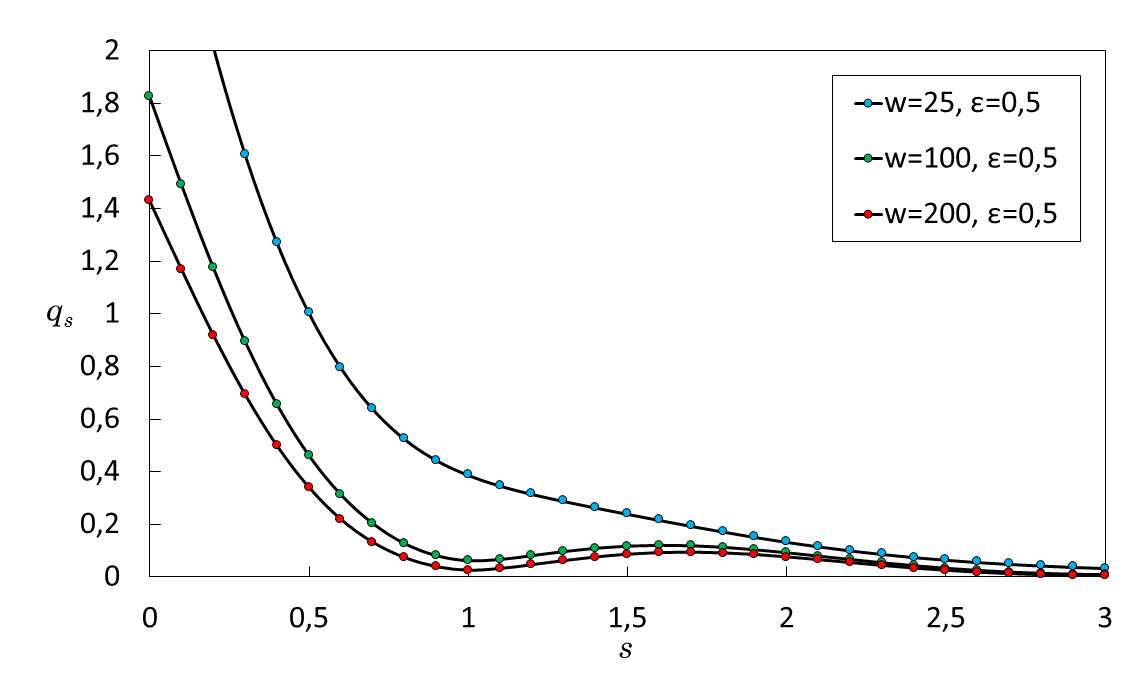}
\end{minipage}%
\begin{minipage}{.5\textwidth}
  \centering
  \includegraphics[scale=0.26]{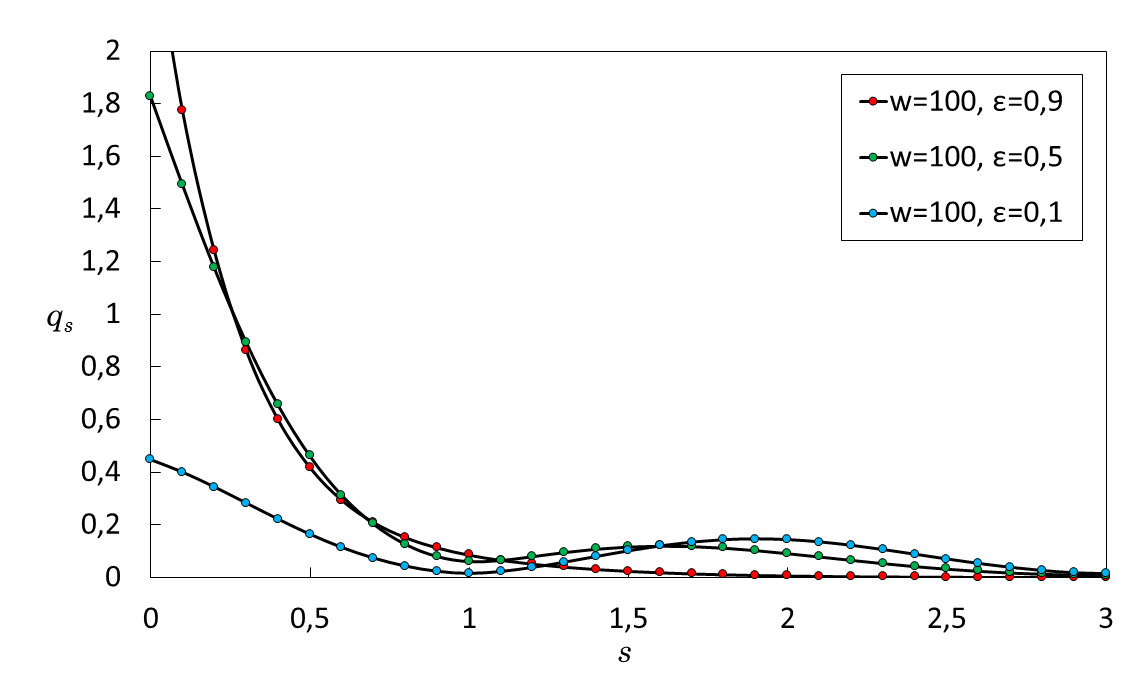}
\end{minipage}
\caption{The figures show the dependence of the variational functions $q_{s}$ on $s$ for different values of the parameters of width $w$ and $\varepsilon$. At the same time, we chose the following system of units: the product $\sqrt{g(0)}D^{2}(0)=1$, as well as the width of the basis functions $\psi_{s}(x)$.}
\label{fig:phi4:qs}
\end{figure}

\begin{figure}[H]
    \centering
    \includegraphics[scale=0.30]{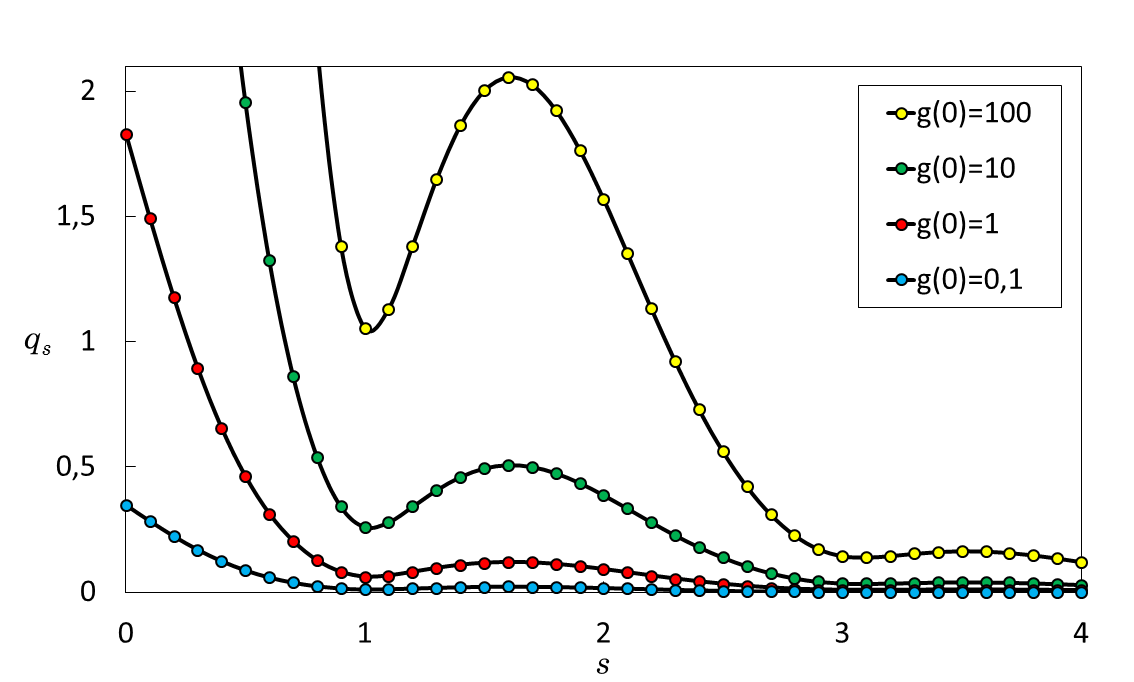}
    \caption{The figure shows the dependence of the functions $q_{s}$ on $s$ for different values of the amplitude of the coupling constant $g(0)$. The parameters $w$ and $\varepsilon$ are fixed and equal to $w=100$ and $\varepsilon=0.5$, respectively. The amplitude of the propagator $D(0)=1$ (in this case, the variation of the product $\sqrt{g(0)}D^{2}(0)$ means the variation of the $g(0)$).}
    \label{fig:phi4:qs_s_g0}
\end{figure}

As one can see from the plots in Fig. \ref{fig:phi4:qs}, \ref{fig:phi4:qs_s_g0} the functions $q_{s}$ decrease rapidly as $s$ grows. Thus, we can cut the summation in the expression for the majorant $F_{\mathrm{sep}}$ (\ref{num:maj}) when calculating it.

\begin{figure}[H]
\centering
\begin{minipage}{.5\textwidth}
  \centering
  \includegraphics[scale=0.26]{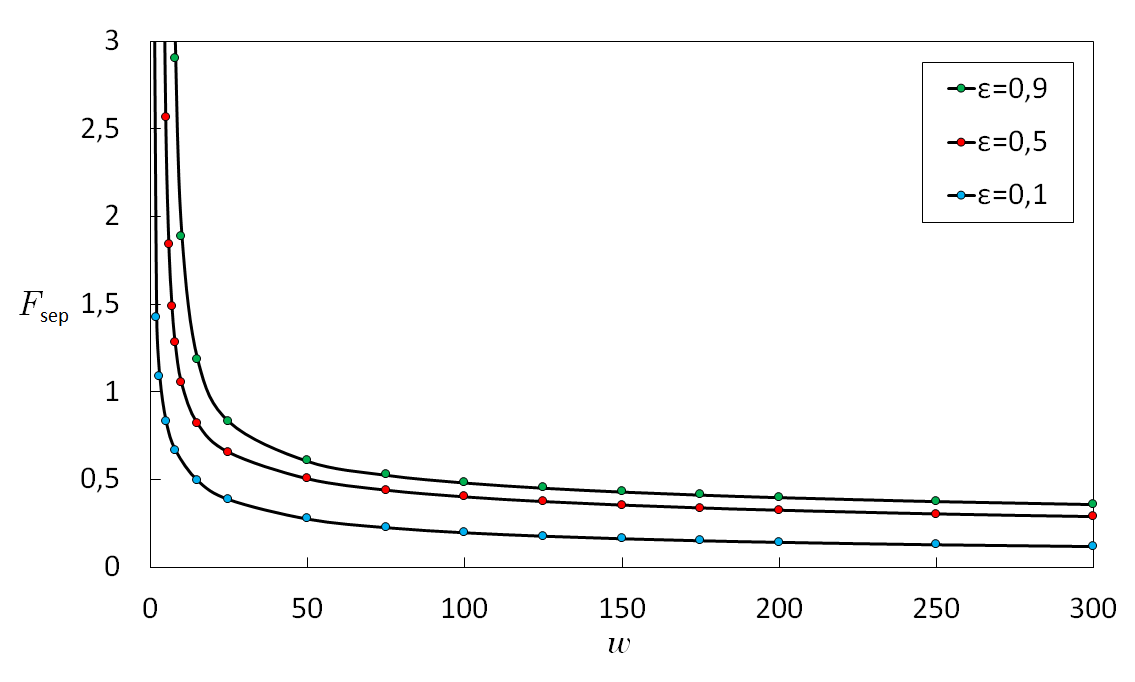}
\end{minipage}%
\begin{minipage}{.5\textwidth}
  \centering
  \includegraphics[scale=0.26]{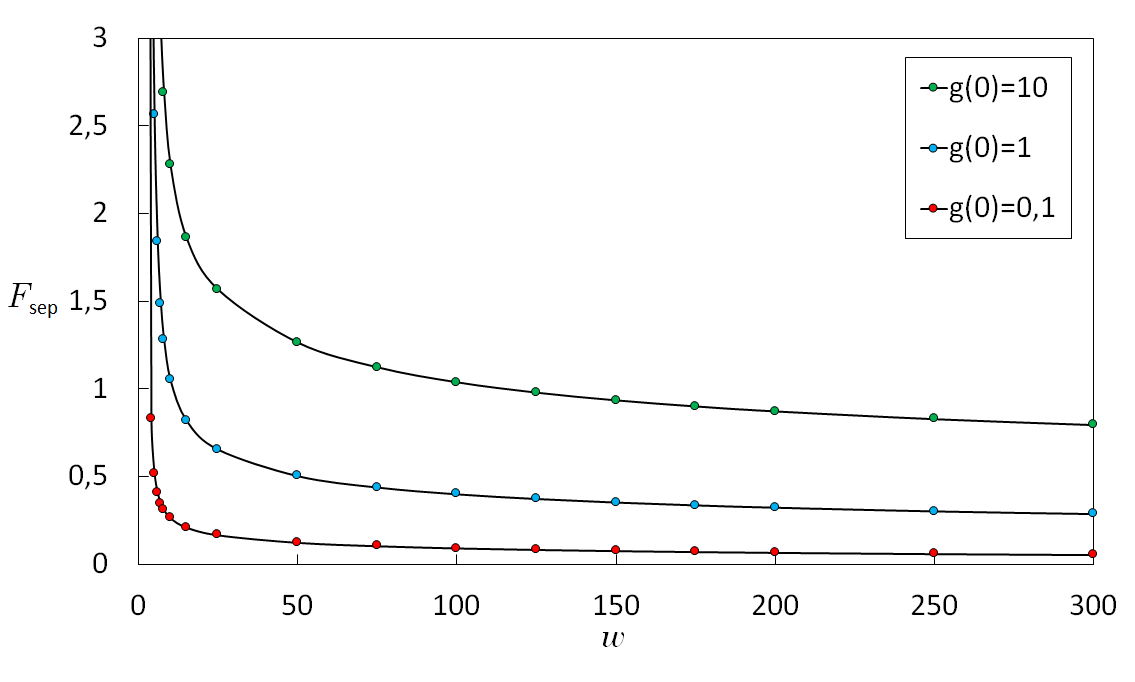}
\end{minipage}
\caption{The figures show the dependence of the majorant $F_{\mathrm{sep}}$ on the width $w$ of the coupling constant $g(x)$ for different values of the parameters $\varepsilon$ and the amplitude of the coupling constant $g(0)$. Left: $\sqrt{g(0)}D^{2}(0)=1$. Right: $\varepsilon=0.5$; $D(0)=1$.}
\label{fig:phi4:maj_w}
\end{figure}

\begin{figure}[H]
    \centering
    \includegraphics[scale=0.30]{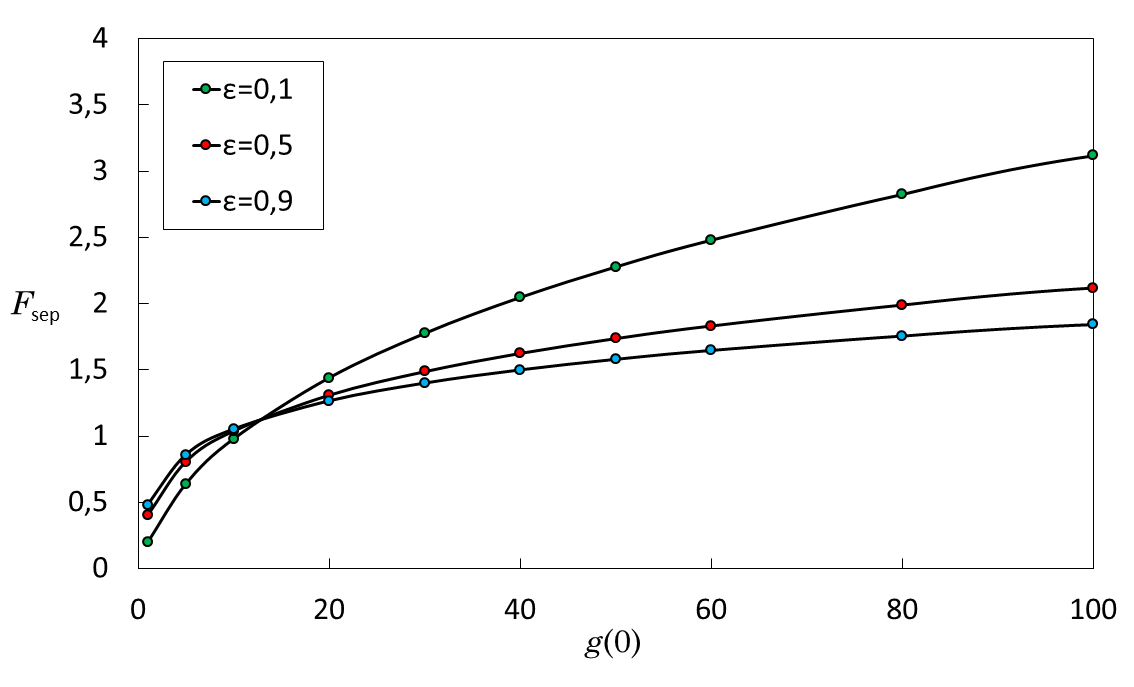}
    \caption{The figure shows the dependence of the majorant $F_{\mathrm{sep}}$ on the amplitude of the coupling constant $g(0)$ for different values of $\varepsilon$. The width parameter $w=100$, the propagator amplitude $D(0)=1$ (which demonstrates the dynamics with respect to the value $g(0)$).}
    \label{fig:phi4:maj_g0_e}
\end{figure}

At the end of the subsection let us make one remark. Selected units system is rather a toy system. These units are designed to demonstrate the existence of the nontrivial solutions within the framework of the general theory developed in the paper. The case $D=1$ is also chosen for simplicity of the numerical calculation, since already in the case of $D=4$ the complexity of the calculation increases significantly. However, in the final subsection we will offer an original point of view on the nonlocal theory in an arbitrary dimension of the spacetime $D$.

\subsection{Nonpolynomial Sine-Gordon Theory}

In this subsection, we consider the majorant $F_{\mathrm{simple}}$ for the $\mathcal{S}$-matrix of the nonpolynomial sine-Gordon theory on the discrete lattice of basis functions. Many values will repeat those of the previous subsection. For this reason, we will focus on the interesting differences that arise in this theory. First consider analytics.

\subsubsection{Analytics}

We begin the analytical calculations with rescaling of the expressions for the upper $q^{(+)}_{s}$ (\ref{sinG_dislat3}) and the lower $q^{(-)}_{s}$ (\ref{sinG_dislat5}) bounds of the variational function $q_{s}$. Then we rescale the expressions for the simplified variational function $q_{s}$ (\ref{sinG_dislat_for_simple5}) and the majorant $F_{\mathrm{simple}}$ (\ref{sinG_dislat_for_numcalc}). Note that this rescaling is possible for any value of the spacetime dimension $D$.

Let us introduce the dimensionless parameter $\alpha$ and rescaled values $\bar{A}^{(+)}$ and $\bar{h}_{s}$ as follows:
\begin{equation}
    \alpha \equiv \eta^{2}G(0), \quad \bar{A}^{(+)} \equiv \frac{A^{(+)}}{G(0)}, \quad \bar{h}_{s} \equiv \frac{h_{s}}{G(0)}.
\label{scaleSG1}
\end{equation}
In terms of (\ref{scaleSG1}), an analytical solution for the upper bound $q^{(+)}_{s}$ (\ref{sinG_dislat3}) reads:
\begin{equation}
    q^{(+)}_{s} = \alpha\bar{A}^{(+)}\bar{h}_{s},\quad \bar{A}^{(+)} = \sum\limits_{s=0}^{\infty}\bar{h}_{s}e^{-\frac{1}{2} \frac{\alpha}{1+\alpha\bar{A}^{(+)}\bar{h}_{s}}}.
\label{scaleSG2}
\end{equation}
Next, we introduce the rescaled values $\bar{A}^{(-)}$ and $\bar{g}^{(0)}_{s}$:
\begin{equation}
    \bar{A}^{(-)} \equiv \frac{A^{(-)}}{G(0)}, \quad \bar{g}^{(0)}_{s} \equiv \frac{g^{(0)}_{s}}{G(0)}.
\label{scaleSG3}
\end{equation}
In terms of (\ref{scaleSG3}), the analytical solution for the lower bound $q^{(-)}_{s}$ (\ref{sinG_dislat5}) reads as follows:
\begin{equation}
    q^{(-)}_{s} = \alpha\bar{g}^{(0)}_{s}e^{-\frac{1}{2}\alpha\bar{A}^{(-)}
    \frac{\bar{h}_{s}}{\bar{g}^{(0)}_{s}}},\quad \bar{A}^{(-)} = \sum\limits_{s=0}^{\infty}\frac{\bar{h}_{s}}{1+ \alpha\bar{g}^{(0)}_{s}e^{-\frac{1}{2}\alpha\bar{A}^{(-)}
    \frac{\bar{h}_{s}}{\bar{g}^{(0)}_{s}}}}.
\label{scaleSG4}
\end{equation}
For solutions (\ref{scaleSG2}) and (\ref{scaleSG4}) in the next subsection, numerical results (plots) will be presented.

Now we consider the simplified variational function $q_{s}$ (\ref{sinG_dislat_for_simple5}) and the majorant $F_{\mathrm{simple}}$ (\ref{sinG_dislat_for_numcalc}). Let us introduce the rescaled value $\bar{A}_{\mathrm{simple}}$:
\begin{equation}
    \bar{A}_{\mathrm{simple}} \equiv \frac{A_{\mathrm{simple}}}{G(0)}.
\label{scaleSG5}
\end{equation}
The analytical solution for the simplified variational function $q_{s}$ (\ref{sinG_dislat_for_simple5}) reads as follows:
\begin{equation}
    q_{s} = \alpha\bar{g}^{(0)}_{s}e^{-\frac{1}{2}
    \frac{\alpha}{g}\bar{A}_{\mathrm{simple}}},\quad \bar{A}_{\mathrm{simple}} = \sum\limits_{s=0}^{\infty} \frac{\bar{g}^{(0)}_{s}}{1+\alpha\bar{g}^{(0)}_{s}  e^{-\frac{1}{2}\frac{\alpha}{g}\bar{A}_{\mathrm{simple}}}}.
\label{scaleSG6}
\end{equation}
Finally, we can write the expression for the majorant $F_{\mathrm{simple}}$ (\ref{sinG_dislat_for_numcalc}) in terms of the values defined in this subsection:
\begin{equation}
    F_{\mathrm{simple}}(q) = \frac{1}{2} \sum\limits_{s=0}^{\infty} \ln{\left(1+\alpha\bar{g}^{(0)}_{s}e^{-\frac{1}{2}
    \frac{\alpha}{g}\bar{A}_{\mathrm{simple}}}\right)} - \left(g+\frac{\alpha\bar{A}_{\mathrm{simple}}}{2}\right)
    e^{-\frac{1}{2}\frac{\alpha}{g}\bar{A}_{\mathrm{simple}}}.
\label{scaleSG7}
\end{equation}
For the solution (\ref{scaleSG6}) and majorant (\ref{scaleSG7}) in the next subsection, numerical results (plots) will also be given. But first we will make an important remark.

The results obtained are determined by the dimensionless parameter $\alpha$. This parameter is the ratio of the ultraviolet length parameter from $G(0)$, and some length parameter from $\eta$. The parameter $\eta$ in turn determines the scale of the interaction Lagrangian (the function $U$, which for the sine-Gordon theory has the form (\ref{for_scale_disc})). The latter will also be considered ultraviolet \cite{basuev1973conv}. A beautiful picture appears: the interaction Lagrangian as a function of the field $U$ is driven by the ultraviolet length scale, but the interaction constant $g(x)$ is driven by some infrared length scale $L$ (through $w$). Such a hierarchy of length scales forms a nontrivial nonpolynomial field theory. Thus, in the future it seems natural to choose the value $\alpha=1$.

\subsubsection{Plots}

In this subsection, we present numerical results (plots) for the majorant $F_{\mathrm{simple}}+g$ of the $\mathcal{S}$-matrix of the nonpolynomial sine-Gordon theory on the discrete lattice of basis functions. The shift $g$ is introduced so that the Lagrangian of the interaction is positive everywhere, which leads to the positivity of the majorant itself by construction (in this case $\mathcal{S}<1$). We also present plots of the variational function $q_{s}$, estimates $q^{(+)}_{s}$ and $q^{(-)}_{s}$. Everywhere accepted $\alpha=1$.

\begin{figure}[H]
\centering
\begin{minipage}{.5\textwidth}
  \centering
  \includegraphics[scale=0.26]{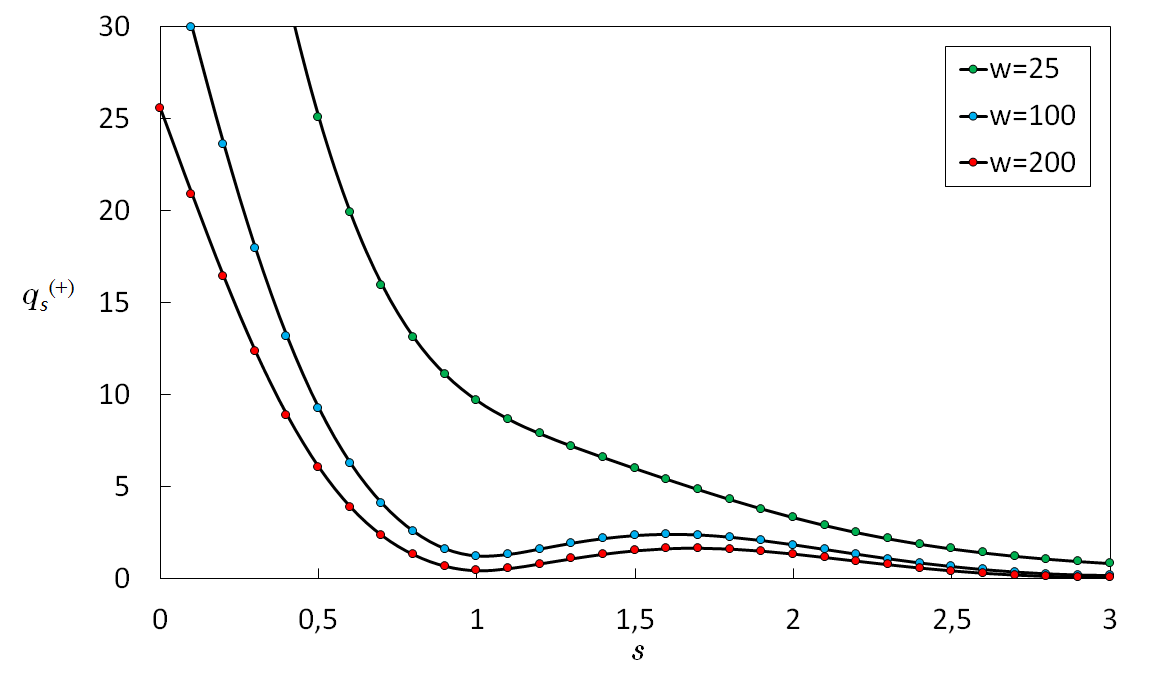}
\end{minipage}%
\begin{minipage}{.5\textwidth}
  \centering
  \includegraphics[scale=0.26]{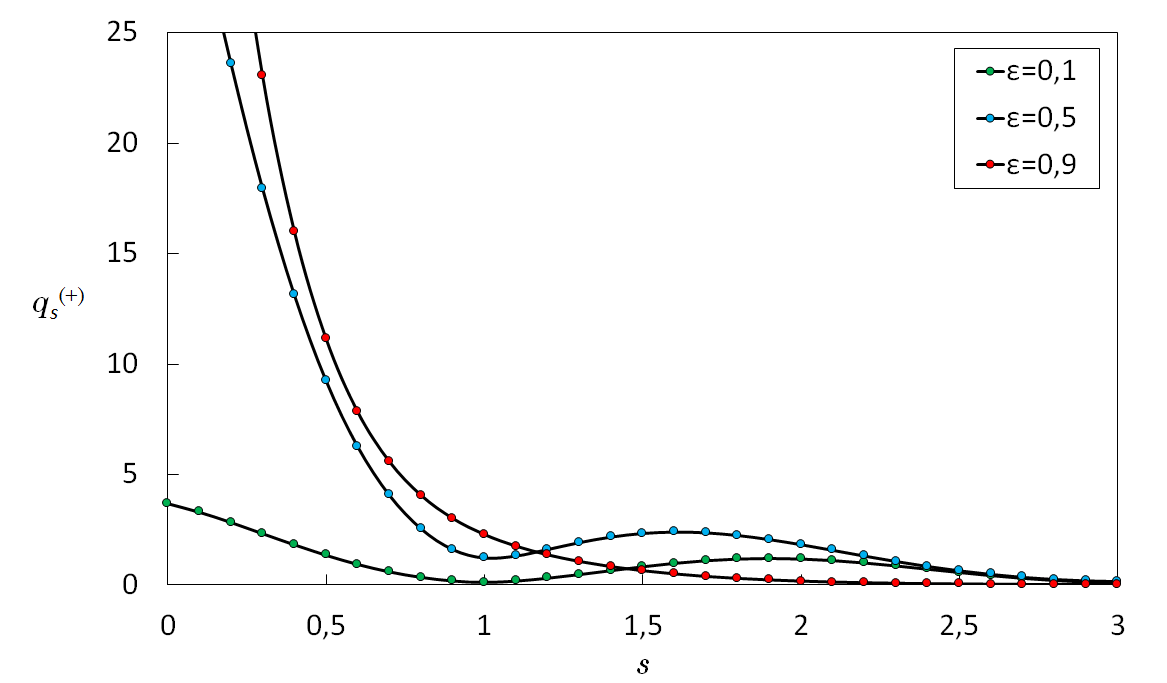}
\end{minipage}
\caption{The figures show the dependence of the variational functions $q^{(+)}_{s}$ on $s$. Left: $w$ changes, $\varepsilon=0.5$. Right: $\varepsilon$ changes, $w=200$. For both plots $g(0)=100$.}
\label{fig:sin_qPlus}
\end{figure}

\begin{figure}[H]
    \centering
    \includegraphics[scale=0.30]{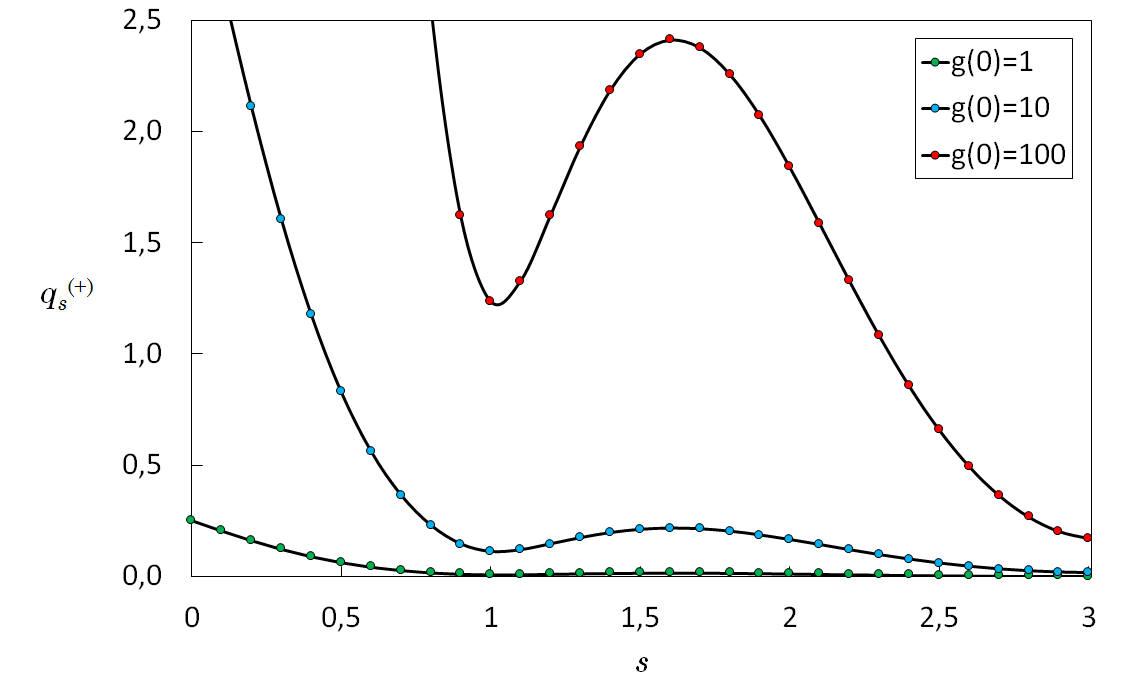}
    \caption{The figure shows the dependence of the functions $q^{(+)}_{s}$ on $s$ for different values of the amplitude of the coupling constant $g(0)$. The parameters $w$ and $\varepsilon$ are fixed and equal to $w=200$ and $\varepsilon=0.5$, respectively.}
    \label{fig:sin_qplusg}
\end{figure}

\begin{figure}[H]
\centering
\begin{minipage}{.5\textwidth}
  \centering
  \includegraphics[scale=0.26]{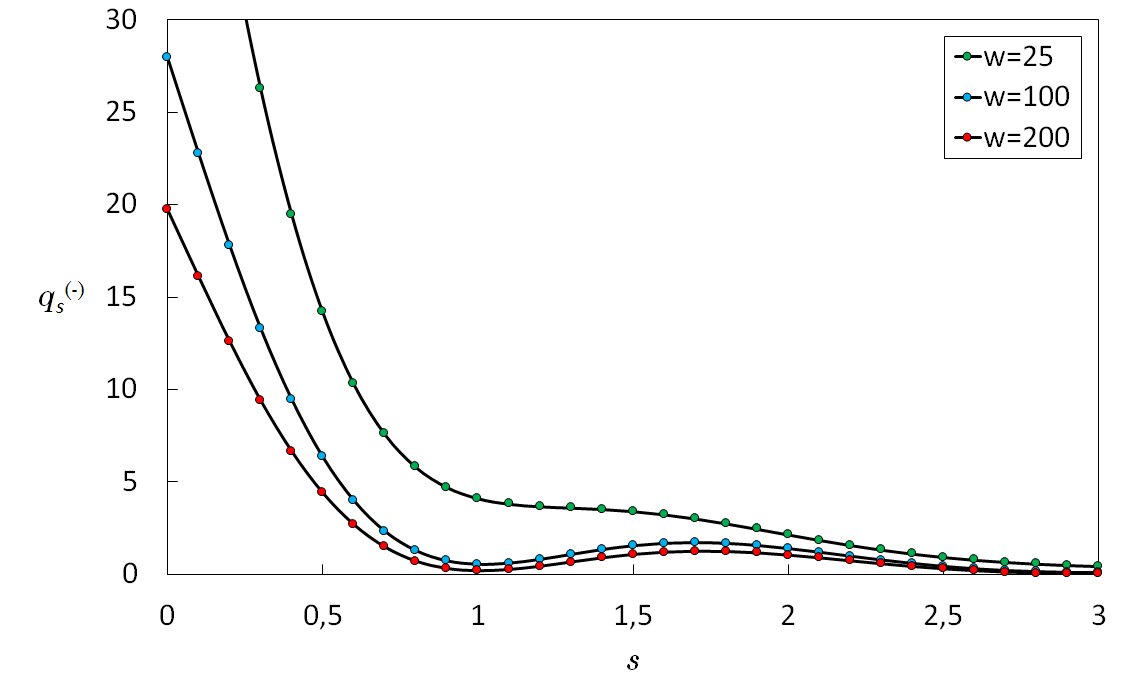}
\end{minipage}%
\begin{minipage}{.5\textwidth}
  \centering
  \includegraphics[scale=0.26]{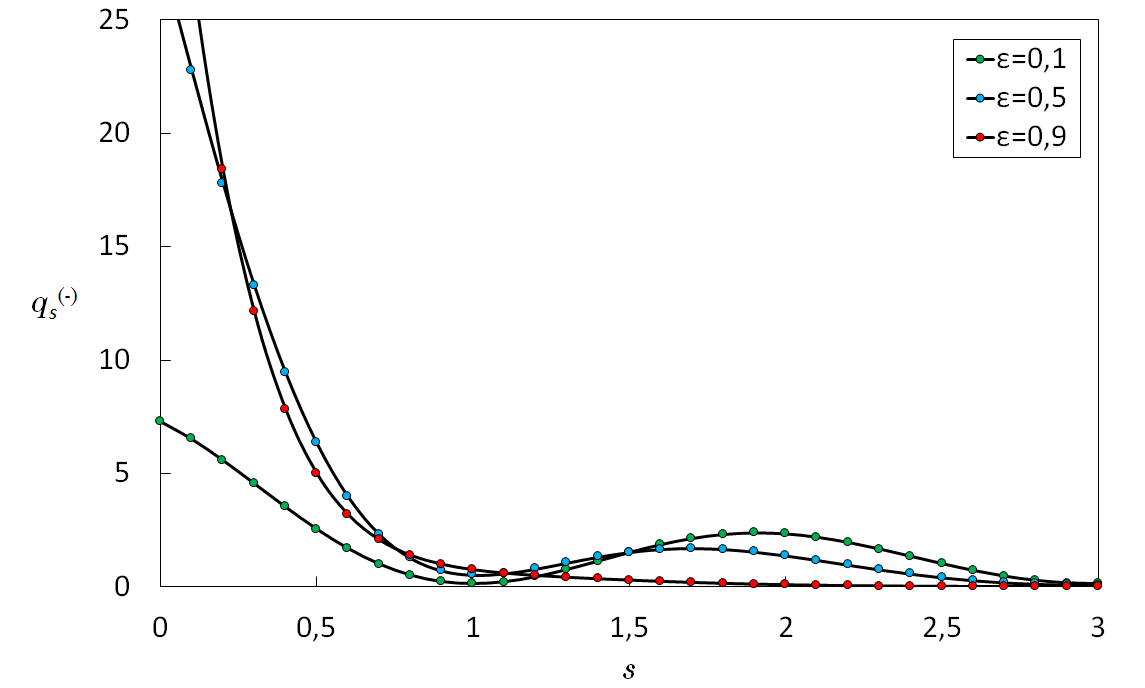}
\end{minipage}
\caption{The figures show the dependence of the variational functions $q^{(-)}_{s}$ on $s$. Left: $w$ changes, $\varepsilon=0.5$. Right: $\varepsilon$ changes, $w=200$. For both plots $g(0)=100$.}
\label{fig:sin_qMinus}
\end{figure}

\begin{figure}[H]
    \centering
    \includegraphics[scale=0.30]{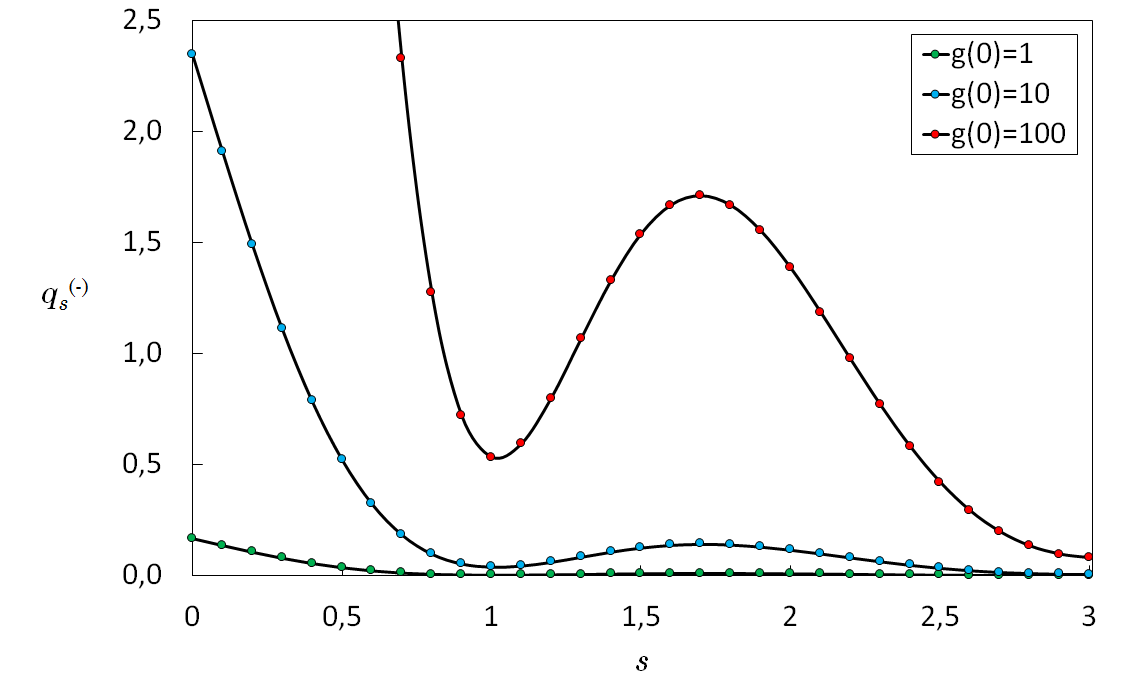}
    \caption{The figure shows the dependence of the functions $q^{(-)}_{s}$ on $s$ for different values of the amplitude of the coupling constant $g(0)$. The parameters $w$ and $\varepsilon$ are fixed and equal to $w=200$ and $\varepsilon=0.5$, respectively.}
    \label{fig:sin_qminusg}
\end{figure}

\begin{figure}[H]
\centering
\begin{minipage}{.5\textwidth}
  \centering
  \includegraphics[scale=0.26]{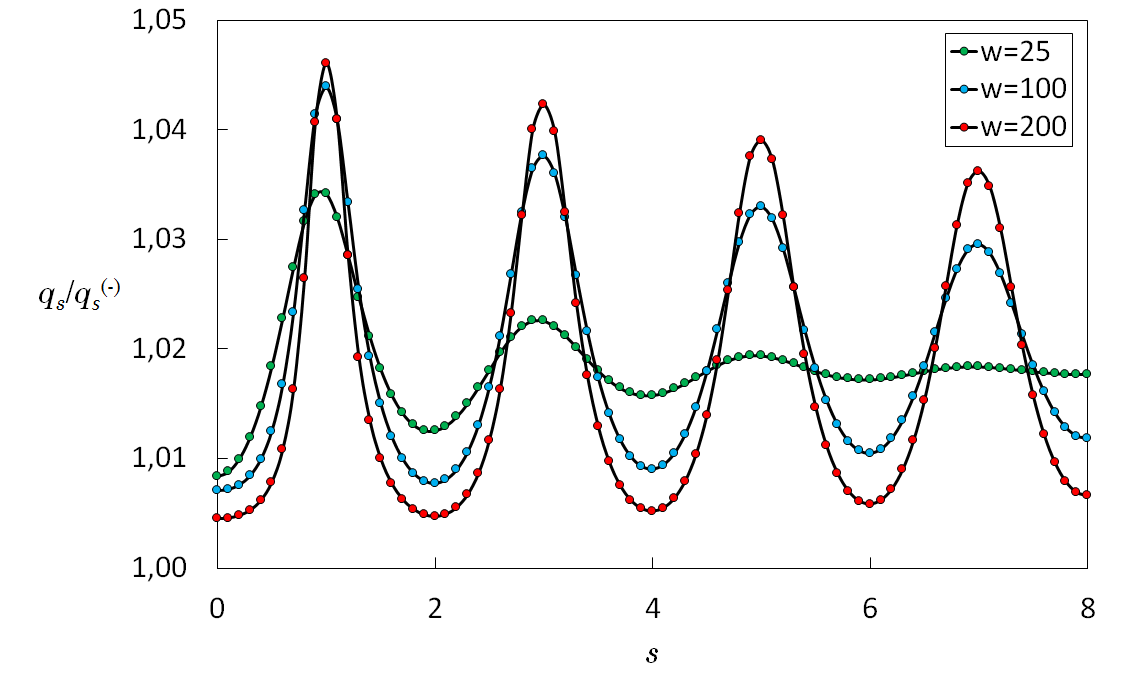}
\end{minipage}%
\begin{minipage}{.5\textwidth}
  \centering
  \includegraphics[scale=0.26]{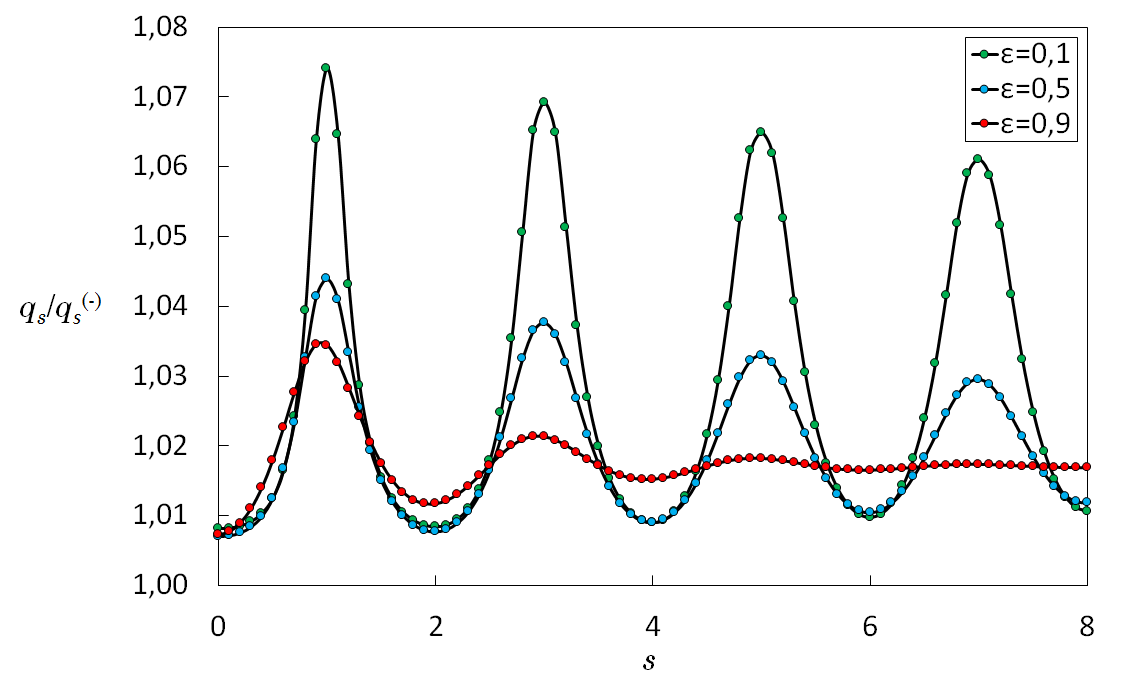}
\end{minipage}
\caption{The figures show the dependence of the ratios $q_{s}/q^{(-)}_{s}$ on $s$. Left: $w$ changes, $\varepsilon=0.5$. Right: $\varepsilon$ changes, $w=200$. For both plots $g(0)=100$.}
\label{fig:sin_qwe}
\end{figure}

\begin{figure}[H]
    \centering
    \includegraphics[scale=0.30]{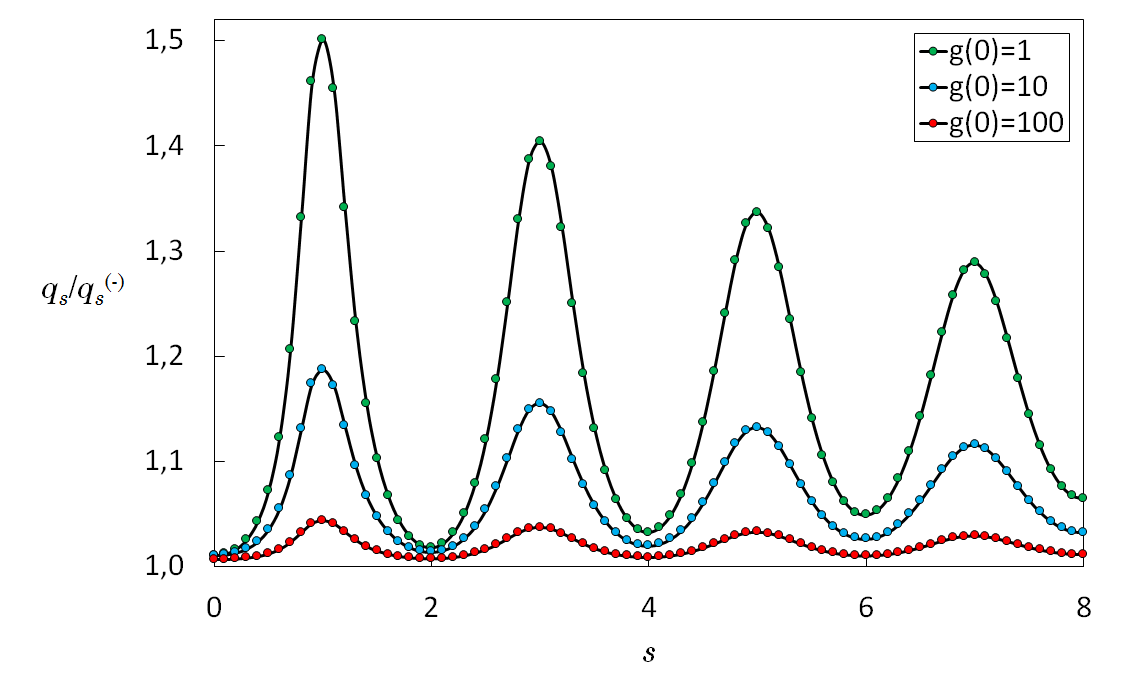}
    \caption{The figure shows the dependence of the ratios $q_{s}/q^{(-)}_{s}$ on $s$ for different values of the amplitude of the coupling constant $g(0)$. The parameters $w$ and $\varepsilon$ are fixed and equal to $w=200$ and $\varepsilon=0.5$, respectively.}
    \label{fig:sin_qg}
\end{figure}

As one can see from the plots in Fig. \ref{fig:sin_qwe}, \ref{fig:sin_qg} simple variational functions $q_{s}$ practically coincide with lower estimates $q^{(-)}_{s}$.

\begin{figure}[H]
\centering
\begin{minipage}{.5\textwidth}
  \centering
  \includegraphics[scale=0.26]{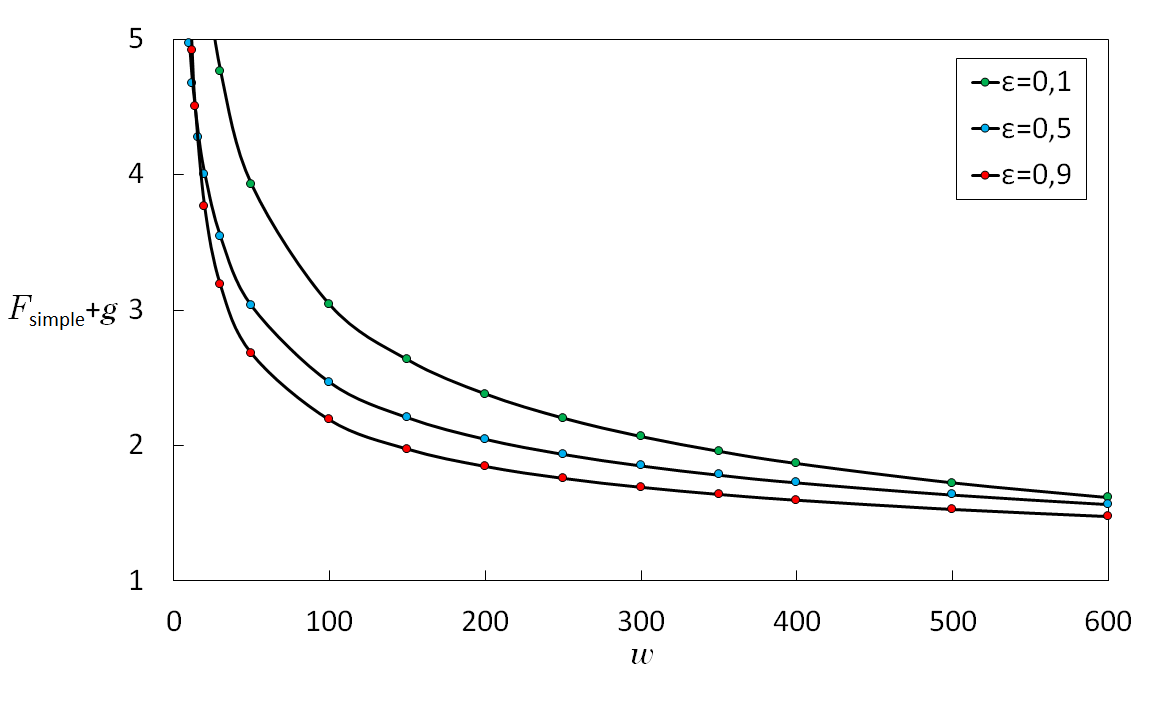}
\end{minipage}%
\begin{minipage}{.5\textwidth}
  \centering
  \includegraphics[scale=0.26]{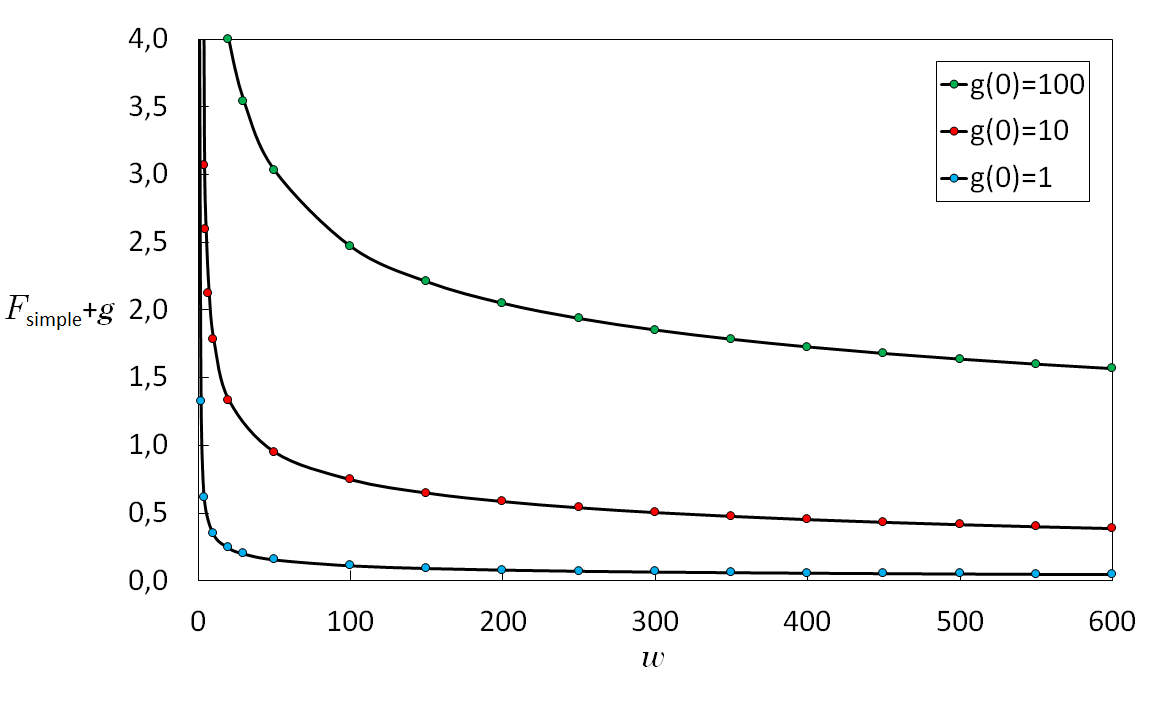}
\end{minipage}
\caption{The figures show the dependence of the majorant $F_{\mathrm{simple}}+g$ on the width of the coupling constant $w$ for different values of the parameters $\varepsilon$ and the amplitude of the coupling constant $g(0)$. Left: $g(0)=100$. Right: $\varepsilon=0.5$.}
\label{fig:sin_Feandg}
\end{figure}

\begin{figure}[H]
    \centering
    \includegraphics[scale=0.30]{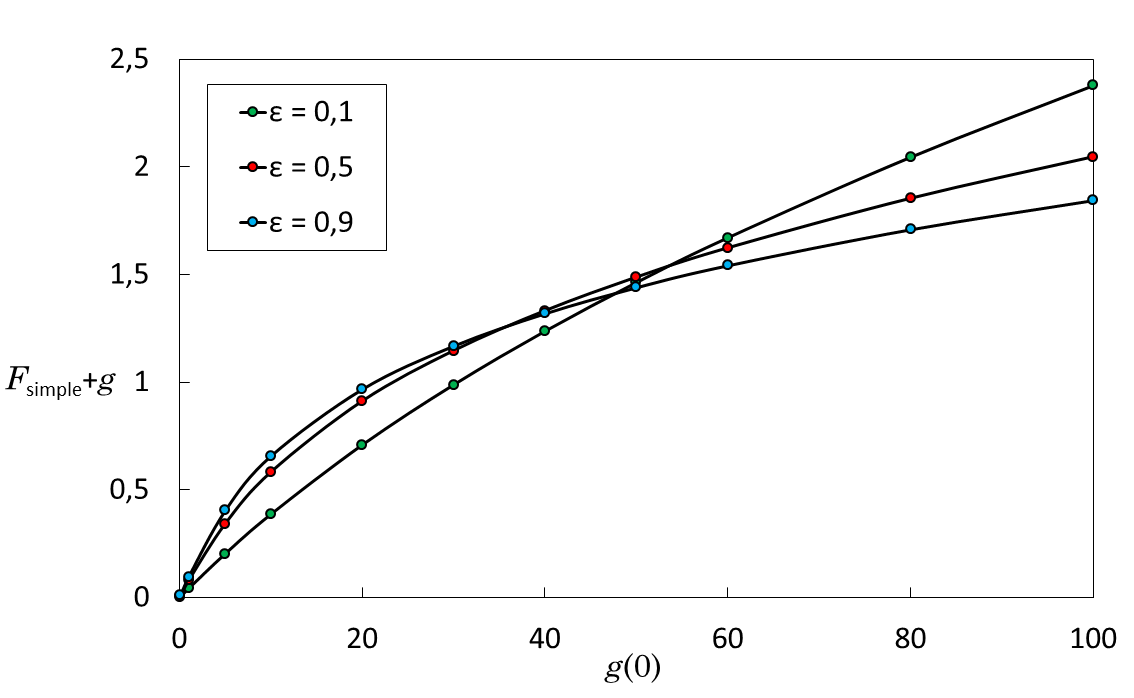}
    \caption{The figure shows the dependence of the majorant $F_{\mathrm{simple}}+g$ on the amplitude of the coupling constant $g(0)$ for different values of the parameter $\varepsilon$. The width parameter $w=200$.}
    \label{fig:sin_Feg}
\end{figure}

At the end of the subsection let us make one remark. Selected units system is already less toy-like, since the choice of $\alpha=1$ describes a realistic nonlocal QFT. However, the case $D=1$ is still chosen for simplicity of the numerical calculation. It is interesting to consider such a theory in the spacetime of high dimension $D$. We discuss the large values of $D$ and their possible role in the nonlocal QFT in the next final subsection.

\subsection{Nonlocal QFT and Compactification Process}

Revisiting expression for propagator $G$ of the free theory in momentum $k$ representation (\ref{PropFexp}): Function $F$ represents ultraviolet form factor, depending on (ultraviolet) length parameter $l$ and infrared mass $m$. Dimension of propagator $G$ is explicitly shown in terms of length parameter $l$, given dimensionless function $\bar{G}$ of dimensionless combinations of arguments. Similar equality is written for exact propagator $\mathcal{G}^{(2)}$ of the  interaction theory (with respect to dimensionless function $\bar{\mathcal{G}}^{(2)}$) as follows:
\begin{equation}
     \tilde{G}(k)=\frac{F(l^{2}k^{2})}{k^{2}+m^{2}}=
     l^{2}\bar{G}(l^{2}k^{2},l^{2}m^{2}),\quad
     \mathcal{G}^{(2)}(k)=l^{2}
     \bar{\mathcal{G}}^{(2)}(l^{2}k^{2},l^{2}m^{2},...).
    \label{Compactif1}
\end{equation}
To obtain a complete picture, the $n$ particle Green function of the theory with interaction $\mathcal{G}^{(n)}$ is represented in similar way (that is, for dimensionless function $\bar{\mathcal{G}}^{(n)}$ such that $D_{n}$ determines dimension of corresponding Green function):
\begin{equation}
     \mathcal{G}^{(n)}(k_{1},...,k_{n})=l^{D_{n}}
     \bar{\mathcal{G}}^{(n)}(lk_{1},...,lk_{n}).
    \label{Compactif2}
\end{equation}
Expressions (\ref{Compactif1}) -- (\ref{Compactif2}) are the subject of the discussion in this subsection.

In framework of the FRG method \cite{kopbarsch} for dimensionless functions $\bar{\mathcal{G}}^{(n)},$ the corresponding flow is constructed over parameter $\varLambda=1/l$, such that $\varLambda$ varies from $\varLambda_{0}$ to $0$. But for every fixed value of $l$ (respectively, a fixed $\varLambda$), it is necessary to understand the role played by $l$ amongst observed physical quantities, moreover, to avoid disagreement with the physical theory which contains the ratio of ultraviolet $l$ and infrared $L$ scales ($L \gg l$). While the presented problem is understandable in the traditional way appealing to different renormalization procedures, we offer a rather original look at what is going on.

Traveling back to the general ideas of the string theory \cite{mohaupt2003,polchinski1998strings,west2016extracoords}, within the framework of this theory, a certain $N$-component $\varphi_{\alpha}$ field is formulated, defined on a two-dimensional manifold $x$, and described by the action of the sigma model. Metrics for both spaces, internal (corresponding to $x$) and external (corresponding to $\varphi_{\alpha}$) are also specified. Since for us the field $\varphi$ is one-component, the metric of corresponding spacetime is trivial; however, there may be nontrivial metric on the manifold $x$, but cannot be two-dimensional from the very beginning. In framework of nonlocal QFT, the action is chosen in arbitrary form, not necessarily in the form of sigma model; for instance, considering the sine-Gordon theory in $D$-dimensional spacetime (we repeat, $D\neq 2$ in the general case). The fields are then expanded according to a basis, and the functional integral is derived by methods of nonlocal QFT framework. As a result, there is a discrepancy with standard string theory.

Moreover, in string theory, low-energy effective action (for instance, action of theory of gravity $+$ gauge field $+$ dilaton) is usually derived in $D$-dimensional spacetime first. The building blocks of the theory are then redefined in terms of equivalents, for instance, gravity constant using dilaton vacuum field value. And, finally, classical field configurations, for example, branes, are introduced into the theory, and/or a \emph{compactification} of $D$-dimensional spacetime, depending on a \emph{process}.

However, we propose the compactification of $D$-dimensional space of momentum $k$ right after deriving the Green functions in terms of the functional integral. In particular, we propose compactification of the $k$ variable in expressions (\ref{Compactif1}) -- (\ref{Compactif2}) right after explicitly deriving the both expressions. That is, for instance, the product $l^{2}k^{2}$ changes into $l_{eff}^{2}k_{4}^{2}$ which corresponds to the four-dimensional spacetime $x_{4}$; at the same time, $l$ and $l_{eff}$ can be quite different. Notably, some QFT models proposed in middle of last century violated the Lorentz invariance of the theory, for instance; they were found to have contained a nontrivial metric of momentum $k_{4}$ space \cite{blokhintsev1973SpaceTime}. Using the nontrivial metric of the internal momentum $k$ space, the problem is avoided. Therefore as a metric, the Euclidean anti-de Sitter space metric is recommended, for instance.

Moreover in further development of the theory, exponential behavior of the Green functions, scattering amplitudes, and form factors, for instance, can be modified into power law or other suitable technique, changing the analytical properties of functions. At the same time, the initial theory (to begin with) satisfies all the requirements for nonlocal QFT. In addition, viewing parameters $l$ and $k$ as some internal variables corresponding to when physical variables changes to $l_{eff}$ and $k_{4}$, brings about a different concept of nonlocal QFT; the theory of nonlocal interactions in the certain internal space has enormous degrees of freedom to describe the physical world; including attempting to construct noncommutative QFT \cite{blokhintsev1973SpaceTime,namsrai2003noncommutative,douglas2001noncommutative}, since the theory is in a sense the field theory with the form factor. Finally, the idea of compactification can be useful even for such exotic field theories as Combinatorial QFTs \cite{reshetikhin2015}, $p$-adic QFTs and $p$-adic strings \cite{volovich1987}.

In summary, actual redefinition of concepts of original theory is proposed. Nonlocal QFT, noncommutative QFT, like other fascinating mathematical constructions of the mid twentieth century, can play worthy role in description of nature, if interpretation of original concepts is changed alongside non-applicability of original building blocks. There is plan to devote separate publication to study of nonlocal QFT in a certain internal space with subsequent compactification for the physical world.
\section{Conclusion}

In this paper, presented in the spirit of others \cite{efimov1970nonlocal,efimov1977nonlocal,efimov1985problems,efimov1977cmp,efimov1979cmp}, we have studied different representations for the $\mathcal{S}$-matrix of nonlocal scalar QFT in Euclidean metric, both for polynomial and nonpolynomial interaction Lagrangians. The theory is formulated on the $D$-dimensional coordinate $x$ and momentum $k$ representations, with the interaction constant $g$ chosen as an infrared-smooth function of its argument such that for large values of the spatial coordinate $x$, the interaction Lagrangian vanishes.

In studying different representations for the $\mathcal{S}$-matrix of the theory, we gave an expression for the initial representation of $\mathcal{S}$ (with respect to the generating functional $\mathcal{Z}$) in terms of abstract functional integral over a primary scalar field. The representation for $\mathcal{S}$ (with respect to $\mathcal{Z}$), in terms of the grand canonical partition function, is then derived from the expression. As indicated in the paper, the grand canonical partition function is a classical series over the interaction constant in the case of a nonpolynomial theory, and at the same time, an abstract (formal) series over the interaction constant in the case of a polynomial QFT. Notably, this derivation is original in the literature.

From the expression for the $\mathcal{S}$-matrix in terms of the grand canonical partition function, the representation for $\mathcal{S}$ in terms of the basis functions is derived. This derivation is presented twice: first for the case of discrete basis functions, which are Hermite functions on discrete lattice of functions, and then for the case of continuous basis functions, which are trigonometric functions on continuous lattice of functions. Notably, the latter is always made square-integrable by a modulating function, but this part is omitted in all expressions of this paper. Moreover all obtained representations for the $\mathcal{S}$-matrix of the theory on discrete and continuous lattices of functions are original.

Obtained expressions for the $\mathcal{S}$-matrix were then investigated within framework of variational principle, based on Jensen inequality. That is, majorant of the corresponding lattice integrals were constructed. Equations with separable kernels satisfied by the variational function $q$ were found and solved, yielding the general theory framework results, that were later studied in detail, both in the case of polynomial theory $\varphi^{4}$ (with suggestions for $\varphi^{6}$) and in the case of the nonpolynomial sine-Gordon theory in $D$-dimensional spacetime. In the latter case, a proposal was made for improving the corresponding majorant. All results described are original.

Moreover, we note that, in applying Jensen inequality in the continuous case, additional divergences arise, from internal structure of the Gaussian integral measure $D\sigma$ on continuous lattice of variables. To solve the problem, we drew parallel to the FRG theory, in particular, we note that the FRG flow regulator $R_{\varLambda}$ contains the measure $d\mu$ as its argument. We therefore, formulated a new definition for the $\mathcal{S}$-matrix of the theory, and corresponding derivations. In addition, under the newly formulated, the obtained results turned out to be consistent in both the discrete and continuous cases of lattice of basis functions, for both polynomial and nonpolynomial QFTs. Yet, the theory turned out to be mathematically rigorous and closed.

Analytical results obtained (on discrete lattice of functions) were illustrated numerically. Plots of variational functions $q$ and corresponding majorants for $\mathcal{S}$-matrices (for $-\ln{\mathcal{S}}$) of scalar nonlocal QFTs discussed in this paper were constructed. For numerical simplicity, we restricted to the case of $D=1$; however, the analytical results obtained do not require the choice to be $D=1$, in principle. Moreover, for numerical simplicity also, all numerical results were obtained for the case where the propagator of the free theory $G$ is a Gaussian function, which is typical in the Virton-Quark model.

An original proposal was made to reinterpret the concept of original nonlocal QFT. We suggested to first consider the theory in $D$-dimensional momentum space $k$, a dimension which exceeds (perhaps, even significantly) the dimension of the physical spacetime. We propose the compactification of such $D$-dimensional space into four-dimensional or below, upon obtaining the Green functions, the scattering amplitudes, and correlation functions of composite operators in terms of the functional integral. For a metric of the original $k$ space, the metric of the Euclidean anti-de Sitter space is recommended, for instance. The proposal mirrors those of disagreement with the ultraviolet and infrared parameters ratio in the theory. Moreover, the proposal makes it possible to change the analytical properties of the Green functions, for instance, if expressed in terms of the compactified (physical) space, where the method of compactification is determined by a process.

To direct further research directions, we note that it is important to consider different methods to improve the variational principle (increasing the accuracy of the majorant), as is a central theme in this paper. We deem it a rightful share to investigate the different generating functionals, the different families of Green functions, and the different composite operators, for new techniques and formulations. Furthermore, in the new direction, there ought to be quest for investigating results obtained in nonlocal QFT framework, for dependency on the ultraviolet parameter $l$. 

An important but challenging research task is the investigation of construction of analytical continuations to Minkowski spacetime, for results obtained in Euclidean nonlocal QFT framework. Another important research area is study of different functional equations associated with the $\mathcal{S}$-matrix of nonlocal QFT, for instance, the Schwinger--Dyson and Tomonaga--Schwinger equations. The study of inclusions of fermions in nonlocal QFT is also important, with the most important model in this direction being the Yukawa model. There is a plan to devote a separate publication to study of the latter model, in the nearest future. Finally, it is important to investigate the nonlocal quantum theory for scalar field in curved spacetime; the simplest instance of this is the Euclidean anti-de Sitter spacetime; a study on this subject is also planned for future publication.

In conclusion, results obtained in this paper to further directions in nonlocal QFT have been made possible, thanks to Gariy Vladimirovich Efimov who laid such a fundamental contribution in the formation and development of the theory in question. To the extent that we were inspired to help advance the theory initiated by G.V. Efimov, in giving our token of research originality, we only prey upon our credulity without G.V. Efimov grappling with the invisible surface. It is only in the combination of G.V. Efimov's scientific works which ended with publications \cite{efimov2007bethesalpeter,efimov2010boundstates,efimov2014scattering,efimov2015particle}, the nontrivial physical intuition with rigorous mathematical closure, that every point of nonlocal QFT methods described and discussed in this paper is able to take a worthy space in description of nature.

\vspace{6pt} 
\authorcontributions{Ivan V. Chebotarev: software, validation, formal analysis, investigation, writing--original draft preparation, visualization; Vladislav A. Guskov: software, validation, formal analysis, investigation, writing--original draft preparation, visualization; Stanislav L. Ogarkov: conceptualization, methodology, formal analysis, investigation, writing--original draft preparation, writing--review and editing, supervision; Matthew Bernard: conceptualization, methodology, formal analysis, investigation, writing--original draft preparation, writing--review and editing.}

\funding{This research received no external funding.}

\acknowledgments{Author S.L. Ogarkov is deeply grateful to his wife Yulia, whose love and wisdom have been the significant support for him throughout the entire research. Author M. Bernard was supported at MIPT by Russian Federation Government with state recommendation of the Ministry of Education and Science of Russian Federation, and is grateful to N.Yu. Reshetikhin for stimulating tasks and conversations on quantum integrability. Authors are grateful to M.A. Ivanov, A. Krikun, S.N. Nedelko and D.B. Rogozkin for fruitful discussions, and are especially grateful to M.G. Ivanov for the detailed discussion of the Hilbert spaces and countable bases theory as well as thorough reading of the manuscript. We express special gratitude to S.E. Kuratov and A.V. Andriyash for the opportunity to develop this research at the Center for Fundamental and Applied Research VNIIA. Finally, the concept of the paper was discussed with J.W. Moffat, to whom the authors are very grateful.}

\abbreviations{The following abbreviations are used in this manuscript:\\
\noindent 
\begin{tabular}{@{}ll}
QFT & Quantum Field Theory\\
QED & Quantum Electrodynamics\\
QCD & Quantum Chromodynamics\\
SM & Standard Model\\
RG & Renormalization Group\\
FRG & Functional Renormalization Group
\end{tabular}}

\reftitle{References}


\end{document}